\pdfoutput=1 %
\documentclass[acmsmall,screen]{acmart}

\setcopyright{rightsretained}
\acmPrice{}
\acmDOI{10.1145/3290377}
\acmYear{2019}
\copyrightyear{2019}
\acmJournal{PACMPL}
\acmVolume{3}
\acmNumber{POPL}
\acmArticle{64}
\acmMonth{1}

\usepackage{booktabs}   %
\usepackage{subcaption} %

\makeatletter%
\@ifundefined{basedir}{%
\newcommand\basedir{}%
}{}%
\makeatother%
\usepackage{amsfonts}
\usepackage{amssymb}
\usepackage{amsthm}
\usepackage{stmaryrd}
\usepackage{multirow}
\usepackage{tabularx}
\usepackage{listingsutf8}
\usepackage{subcaption}
\usepackage{breakcites}

\usepackage{\basedir mathpartir}

\usepackage{\basedir pftools}
\usepackage{\basedir iris}

\usepackage{xcolor}  %

\usepackage{graphicx}
\usepackage{enumitem}
\usepackage{semantic}
\usepackage{csquotes}
\usepackage{letltxmacro}

\usepackage{hyperref}

\SetSymbolFont{stmry}{bold}{U}{stmry}{m}{n} %

\extrarowheight=\jot	%
\newcolumntype{.}{@{}}
\newcolumntype{M}{@{\mskip\thickmuskip}}

\definecolor{StringRed}{rgb}{.637,0.082,0.082}
\definecolor{CommentGreen}{rgb}{0.0,0.55,0.3}
\definecolor{KeywordBlue}{rgb}{0.0,0.3,0.55}
\definecolor{LinkColor}{rgb}{0.55,0.0,0.3}
\definecolor{CiteColor}{rgb}{0.55,0.0,0.3}
\definecolor{HighlightColor}{rgb}{0.0,0.0,0.0}

\definecolor{grey}{rgb}{0.5,0.5,0.5}
\definecolor{red}{rgb}{1,0,0}

\hypersetup{%
  linktocpage=true, pdfstartview=FitV,
  breaklinks=true, pageanchor=true, pdfpagemode=UseOutlines,
  plainpages=false, bookmarksnumbered, bookmarksopen=true, bookmarksopenlevel=3,
  hypertexnames=true, pdfhighlight=/O,
  colorlinks=true,linkcolor=LinkColor,citecolor=CiteColor,
  urlcolor=LinkColor
}

\newcommand*{\Sref}[1]{\hyperref[#1]{\S\ref*{#1}}}

\newcommand*{\lemref}[1]{\hyperref[#1]{Lemma~\ref*{#1}}}
\newcommand*{\thmref}[1]{\hyperref[#1]{Theorem~\ref*{#1}}}
\newcommand{\corref}[1]{\hyperref[#1]{Cor.~\ref*{#1}}}
\newcommand*{\defref}[1]{\hyperref[#1]{Definition~\ref*{#1}}}
\newcommand*{\egref}[1]{\hyperref[#1]{Example~\ref*{#1}}}
\newcommand*{\figref}[1]{\hyperref[#1]{Figure~\ref*{#1}}}
\newcommand*{\tabref}[1]{\hyperref[#1]{Table~\ref*{#1}}}

\newcommand{\ie}{\emph{i.e.,} }
\newcommand{\eg}{\emph{e.g.,} }

\renewcommand*{\mathellipsis}{%
  \mathinner{{\ldotp}{\ldotp}{\ldotp}}%
}
\makeatletter
\@ifdefinable{\org@ldots}{%
  \LetLtxMacro\org@ldots\ldots
  \DeclareRobustCommand*{\ldots}{%
    \ifmmode
      \expandafter\my@ldots
    \else
      \expandafter\textellipsis
    \fi
  }%
}
\newcommand*{\neghalfmskip}{%
  \nonscript\mskip-.5\muexpr\thinmuskip\relax%
}
\newcommand*{\my@ldots}{%
  \mathellipsis
  \@ifnextchar,\neghalfmskip{%
  \@ifnextchar:\neghalfmskip{%
  \@ifnextchar;\neghalfmskip{%
  \@ifnextchar.\neghalfmskip{%
  \@ifnextchar!\neghalfmskip{%
  \@ifnextchar?\neghalfmskip{%
    \rightdelim@
    \ifgtest@
      \mskip-.5\muexpr\thinmuskip\relax%
    \fi
  }}}}}}%
}
\makeatother

\newcommand{\loc}{l}

\newcommand{\vunit}{()}

\newcommand{\fork}[1]{\textlog{fork}\{#1\}}

\newcommand{\bind}[3]{\textlog{let}\;#1\;\textlog{=}\;#2\;\textlog{in}\;#3}
\newcommand{\flip}[1]{\textlog{flip}(#1)}
\newcommand{\flipNoArgs}{\textlog{flip}}
\newcommand{\randbits}[1]{\textlog{randbits}(#1)}

\newcommand{\store}[2]{#1\;\textlog{:=}\;#2}
\newcommand{\match}[1]{\textlog{match}\;#1\;\textlog{with}}
\newcommand{\ifmatch}[1]{\textlog{if}\;#1\;\textlog{then}}
\newcommand{\elsematch}{\textlog{else}}
\newcommand{\matchend}{\textlog{end}}
\newcommand{\load}[1]{! #1}
\newcommand{\alloc}[1]{\textlog{ref}\;#1}

\newcommand{\inl}{\textlog{inl}}
\newcommand{\inr}{\textlog{inr}}

\let\mapsto\hookrightarrow

\newcommand\Val{V}

\newcommand{\ownProb}[1]{\textlog{Prob}(#1)}
\newcommand{\ownProbNoArgs}{\textlog{Prob}}

\makeatletter %
\def\arcr{\@arraycr}
\makeatother

\newcommand{\isnew}[1]{{\color{blue} #1}}

\newtheorem*{thesis}{Thesis}

\newcommand{\sched}{\varphi}
\newcommand{\trstep}[2]{\xrightarrow[#2]{#1}}
\newcommand{\trstepival}[2]{\textlog{tstep}_{#1}(#2)}
\newcommand{\trstepivalN}[3]{\textlog{resStep}^{#3}_{#1}(#2)}

\newcommand{\trace}{T}
\newcommand{\curr}{\textlog{curr}}

\newcommand{\probmonad}[1]{M_{\textsf{P}}(#1)}
\newcommand{\probmonadNoArg}{M_{\textsf{P}}}
\newcommand{\nondetmonad}[1]{M_{\textsf{N}}(#1)}
\newcommand{\nondetmonadNoArg}{M_{\textsf{N}}}
\newcommand{\ivalmonad}[1]{M_{\textsf{I}}(#1)}
\newcommand{\ivalmonadNoArg}{M_{\textsf{I}}}
\newcommand{\pivalmonad}[1]{M_{\textsf{NI}}(#1)}
\newcommand{\pivalmonadNoArg}{M_{\textsf{NI}}}

\newcommand{\ndchoice}{\cup}
\newcommand{\pchoice}[1]{\oplus_{#1}}

\newcommand{\incrfun}{\textlog{incr}}
\newcommand{\incrauxfun}{\textlog{incr\_aux}}
\newcommand{\readfun}{\textlog{read}}
\newcommand{\compareSwap}[3]{\textlog{CAS}(#1, #2, #3)}
\newcommand{\compareSwapNoArg}{\textlog{CAS}}
\newcommand{\fetchAdd}[2]{\textlog{FAA}(#1, #2)}
\newcommand{\fetchAddNoArg}{\textlog{FAA}}
\newcommand{\lsbZero}[2]{\textlog{lsbZero}(#1, #2)}
\newcommand{\MAXCONST}{\textlog{MAX}}
\newcommand{\mincmd}[2]{\textlog{min}(#1, #2)}

\newcommand{\Skipprop}[6]{\textlog{SkipPerm}_{#1}(#2, #3, #4, #5, #6)}
\newcommand{\SkippropNoArgs}{\textlog{SkipPerm}}
\newcommand{\skipPival}{\textlog{skiplist}}
\newcommand{\addfun}{\textlog{addSkipList}}

\newcommand{\memfun}{\textlog{memSkipList}}
\newcommand{\newslfun}{\textlog{newSkipList}}

\newcommand{\nil}{\textlog{nil}}
\newcommand{\Gbundle}{\Gamma}
\newcommand{\INTMIN}{\textlog{INTMIN}}
\newcommand{\INTMAX}{\textlog{INTMAX}}
\newcommand{\skipcost}[3]{\textlog{skipcost}(#1, #2, #3)}
\newcommand{\skipcostNoArg}{\textlog{skipcost}}
\newcommand{\topcost}[2]{\textlog{topcost}(#1, #2)}
\newcommand{\botcost}[3]{\textlog{botcost}(#1, #2, #3)}
\newcommand{\sort}{\textlog{sort}}
\newcommand{\removeList}{\textlog{remove}}
\newcommand{\rettop}[2]{\textlog{maxbelow}(#1, #2)}

\newcommand{\pival}{\mathcal{I}}
\newcommand{\ival}{\mathbb{I}}
\newcommand{\idxset}{I}
\newcommand{\valfun}{\textlog{val}}
\newcommand{\indfun}{\textlog{ind}}
\newcommand{\ivaltoprob}{H}

\newcommand{\expected}[1]{\mathbb{E}[#1]}
\newcommand{\exival}[2]{\mathbb{E}_{#1}[#2]}
\newcommand{\exMax}[2]{\mathbb{E}^{\textlog{max}}_{#1}[#2]}
\newcommand{\exMaxNoArgs}{\mathbb{E}^{\textlog{max}}}
\newcommand{\exMin}[2]{\mathbb{E}^{\textlog{min}}_{#1}[#2]}

\newcommand{\exMinBig}[2]{\mathbb{E}^{\textlog{min}}_{#1}\left[#2\right]}

\newcommand{\mbind}[3]{#1\leftarrow#2\, ;\;#3}
\newcommand{\mret}[1]{\textlog{ret}\; #1}

\newcommand{\ipCoupling}[3]{#1 \sim #2 : #3}

\newcommand{\pivalleq}{\subseteq}

\newcommand{\irrelequiv}{\equiv_{\textlog{p}}}
\newcommand{\irrelleq}{\subseteq_{\textlog{p}}}

\newcommand{\authFullCount}[1]{\authfull #1}
\newcommand{\authFragCount}[2]{\authfrag(#1, #2)}

\newcommand{\funsupp}[1]{\textlog{supp}(#1)}
\newcommand{\idxsupp}[1]{\textlog{isupp}(#1)}
\newcommand{\Isupp}[1]{\textlog{supp}(#1)}
\newcommand{\Psupp}[1]{\textlog{supp}(#1)}

\newcommand{\PROP}{\textlog{Prop}}

\newcommand*{\INCLUDEAPPENDIX}{}%

\ifdefined\INCLUDEAPPENDIX
\newcommand*{\appendixref}[1]{\hyperref[#1]{Appendix~\ref*{#1}}}
\else
\newcommand*{\appendixref}[1]{the supplemental appendix}
\fi

\begin{document}

\title[A Separation Logic for Concurrent Randomized Programs]{A Separation Logic for Concurrent Randomized Programs}

\author{Joseph Tassarotti}
\affiliation{
  \department{Computer Science Dpartment}              %
  \institution{Carnegie Mellon University}            %
  \country{USA}                    %
}
\email{jtassaro@andrew.cmu.edu}          %

\author{Robert Harper}
\affiliation{
  \department{Computer Science Department}             %
  \institution{Carnegie Mellon University}           %
  \country{USA}                    %
}
\email{rwh@cs.cmu.edu}         %



\begin{abstract}
  We present Polaris, a concurrent separation logic with support for probabilistic reasoning.
  As part of our logic, we extend the idea of \emph{coupling}, which underlies recent work on probabilistic relational logics,
  to the setting of programs with both probabilistic and non-deterministic choice.
  To demonstrate Polaris, we verify a variant of a randomized concurrent counter
  algorithm and a two-level concurrent skip list. All of our results have been
  mechanized in Coq.

\end{abstract}
 
\maketitle

\section{Introduction}
\label{sec:intro}

Many concurrent algorithms use randomization to reduce contention and
coordination between threads. Roughly speaking, these algorithms are
designed so that if each thread makes a local random choice, then on average the
aggregate behavior of the whole system will have some good property.

For example, probabilistic skip lists~\citep{Pugh90} are known to work well in
the concurrent setting~\citep{HLLS-Skip, Fraser04}, because threads can
independently insert nodes into the skip list without much synchronization. In
contrast, traditional balanced tree structures are difficult to implement in a
scalable way because re-balancing operations may require locking access to large
parts of the tree.

However, concurrent randomized algorithms are difficult to write and
reason about.  The use of just concurrency or
randomness alone makes it hard to establish the correctness of an
algorithm. For that reason, a number of program logics for reasoning
about concurrent~\citep{ohearn:csl, rgsep, cap, iris, hlrg, fcsl,
  views, rg} or randomized~\citep{MorganMS96, Ramshaw:1979,
  BartheGGHS16, BartheGB12, KaminskiKMO16} programs have been
developed.

But, to our knowledge, the only prior program logic designed for reasoning about
programs that are both concurrent \emph{and} randomized is the recent
probabilistic rely-guarantee calculus developed by \citet{McIverRS16}, which
extends Jones's original rely-guarantee logic~\citep{rg} with probabilistic
constructs.  However, this logic lacks many of the features of modern
concurrency logics.
For example, starting with the work of \citet{rgsep} and \citet{sagl}, many
recent concurrency logics combine rely-guarantee style reasoning with some form
of separation logic, which is useful for modular, local reasoning about
fine-grained concurrent data structures.

In this paper we describe Polaris, a logic which extends Iris~\citep{iris}, a state of the
art concurrency logic, with support for probabilistic reasoning. By extending Iris,
we ensure that our logic has the features needed to reason modularly
about sophisticated concurrent algorithms. The key to our
approach is several recent developments in probabilistic logic, concurrency
logics, and denotational semantics. However, before we give an overview of how
we build on this related work, let us describe a concurrent randomized
algorithm, which will be our running example throughout this paper.

\subsection{Example: Concurrent Approximate Counters}

\newcommand\counterfig{\begin{figure*}
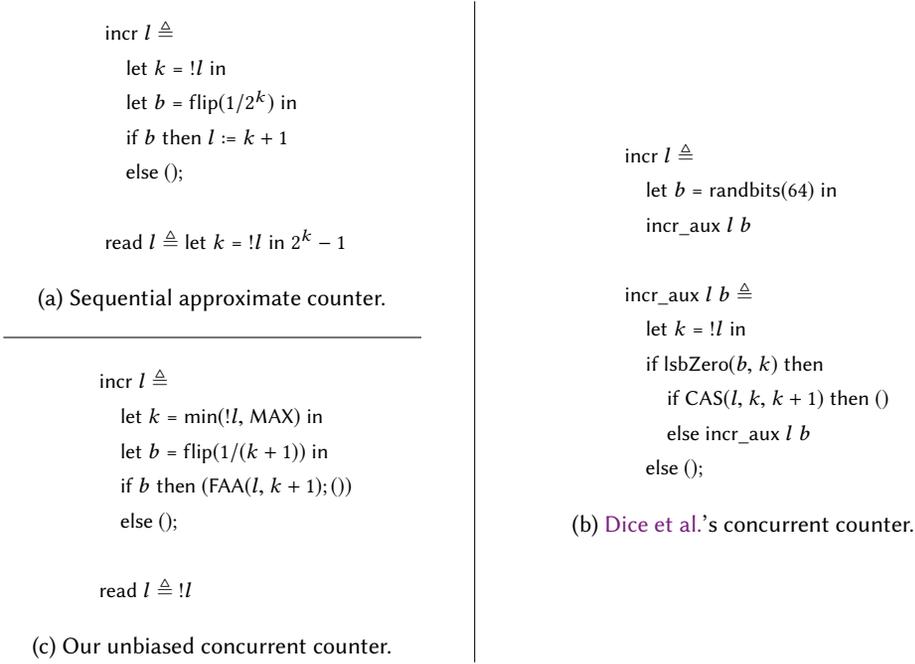

  \centering
  \begin{subfigure}[c]{0.4\textwidth}
    {\footnotesize
    \centering
    \[
    \begin{array}{ll}
      &\incrfun\  l\eqdef  \\
      &\quad \bind{k}{\load{l}}{} \\
      &\quad \bind{b}{\flip{1/2^k}}{} \\
      &\quad \ifmatch{b}\ \store{l}{k+1} \\
      &\quad \elsematch\ ();
      \\ \\
      &\readfun\  l\eqdef \bind{k}{\load{l}}{2^k - 1}
    \end{array}
    \]
    }
    \caption{Sequential approximate counter.}
    \label{fig:morris}
  \vspace{2ex}
  \hrule
  \addtocounter{subfigure}{1}
  \vspace{1ex}
  {\footnotesize
    \centering
    \[
    \begin{array}{ll}
      &\incrfun\  l\eqdef  \\
      &\quad \bind{k}{\mincmd{\load{l}}{\MAXCONST}}{} \\
      &\quad \bind{b}{\flip{1/(k + 1)}}{} \\
      &\quad \ifmatch{b}\ (\fetchAdd{l}{k+1}; ()) \\
      &\quad \elsematch\ ();
      \\ \\ 
      &\readfun\  l\eqdef\ \load{l}
    \end{array}
    \]
    }
    \caption{Our unbiased concurrent counter.}
    \label{fig:unbiased}
  \end{subfigure}%
  \qquad
  \vrule
  \qquad
  \addtocounter{subfigure}{-2}
  \begin{subfigure}[c]{0.4\textwidth}
    {\footnotesize
    \centering
    \[
    \begin{array}{ll}
      &\incrfun\  l\eqdef  \\
      &\quad \bind{b}{\randbits{64}}{} \\
      &\quad \incrauxfun\ l\ b
      \\ \\
      &\incrauxfun\  l\ b\eqdef  \\
      &\quad \bind{k}{\load{l}}{} \\
      &\quad \ifmatch{\lsbZero{b}{k}} \\
      &\quad\quad \ifmatch{\compareSwap{l}{k}{k+1}}\ () \\
      &\quad\quad \elsematch\ \incrauxfun\ l\ b \\
      &\quad \elsematch\ ();
    \end{array}
    \]
    }
    \caption{\citeauthor{DiceLM13}'s concurrent counter.}
    \label{fig:dlm}
  \end{subfigure}%
\caption{Approximate counting algorithms.}
\label{fig:implementation}
\end{figure*}
}
 \counterfig

In many concurrent systems, threads need to keep counts of events. For
example, in OS kernels, these counts can track performance statistics
or reference counts. Somewhat surprisingly, \citet{Boyd-Wickizer10}
have shown that maintaining such counts was a serious scalability
bottleneck in a prior version of the Linux kernel. In many cases,
however, there is no need for these counts to be \emph{exactly} right:
an estimate is good enough. Taking advantage of this,
\citet{DiceLM13} created a scalable concurrent counter by
adapting \citeauthor{Morris78a}'s~\citep{Morris78a} \emph{approximate counting} algorithm.

In order to understand \citeauthor{DiceLM13}'s concurrent version it is helpful
to understand \citeauthor{Morris78a}'s original work.
Morris's
motivation was to be able to count up to $n$ using fewer
than $O(\log_2(n))$ bits. His idea was that, rather than storing
the exact count $n$, one could instead store something like $\log_2(n)$ rounded
to the nearest integer. This would require only $O(\log_2\log_2(n))$ bits,
at the cost of the error introduced by rounding.

Of course, if we round the stored count, then when we need to
increment the counter, we do not know how to update the rounded value
correctly. Instead, \citeauthor{Morris78a} developed a randomized
increment routine: if the counter currently stores the value $k$, then
with probability $\frac{1}{2^k}$ we update the stored value to $k+1$
and otherwise leave it unchanged.  The code for this increment
function is shown in \figref{fig:morris}, written in an ML-like pseudo-code,
where $\flip{p}$ is a command returning $\TRUE$ with probability
$p$ and $\FALSE$ otherwise.  The $\readfun$ function loads the current
value $k$ of the counter and returns $2^k - 1$.  Let $C_n$ be the
random variable giving the value stored in the counter after $n$ calls
of the increment function.  One can show that $\expected{2^{C_n} - 1}
= n$. Thus, the expected value of the result returned by $\readfun$
is equal to the true number of increments, and so the counter is said to be an
\emph{unbiased} estimator. \citet{Flajolet85} gave a very detailed
analysis of this algorithm by showing that it can be modeled by a
simple Markov chain, and proved that it indeed only requires
$O(\log_2\log_2(n))$ bits with high probability.

Although this space-saving property is interesting, the aspect of
the algorithm that makes it useful for concurrent counting is that, as
the stored count gets larger, the probability that an increment
needs to write to memory to update the count gets smaller.
A simplified version of the concurrent algorithm proposed by
\citet{DiceLM13} is shown in \figref{fig:dlm}.  The increment
procedure starts by generating a large number of uniformly random bits. It then
calls a recursive helper function $\incrauxfun$ with the random bit
vector $b$ as an argument. This helper function reads the current value
$k$ stored in the counter and checks whether the first $k$ bits of the
bitvector are all $0$ (performed in the code by the $\lsbZero{k}{b}$
function). If not, the increment is over. If they are all zero, then
because this occurs with probability $\frac{1}{2^k}$, the thread
tries to atomically update the value stored in the counter from $k$ to
$k+1$ using a compare-and-swap (\compareSwapNoArg) operation. If this
operation succeeds, it means that no other thread has intervened and
modified the counter, and so the increment is finished.  If the swap
fails, some other thread has modified the counter, so the
$\incrauxfun$ function is recursively called to try again. The read procedure
is the same as for the sequential algorithm. 

As the count gets larger, the probability that a thread will perform a
\compareSwapNoArg\ operation during the increment gets smaller, which is
useful because these operations are slow.
\citet{DiceLM13} show that this algorithm works quite well in
practice, but do not give a formal argument for its
correctness. Therefore, one might ask whether it is still guaranteed
to give an unbiased estimate of the count. In fact, the answer is
no: the scheduler can bias the count
by ordering the compare-and-swap operations in a particular way. 
Imagine multiple threads are attempting to concurrently
perform an increment, and the scheduler lets them each generate their
random bits and then pauses them.  Suppose the
current value in the counter is $k$.  Some of the threads may have
drawn values that would cause them to not do an
increment, because there is a 1 within their first $k$ bits.
Others may have drawn a number where far more than the first
$k$ bits will be $0$: these threads would have performed an increment
even if the value in the counter were larger than $k$. The scheduler
can exploit this fact to maximize the value of the counter by running
each thread one after the other in order of how many $0$ bits they
have at the beginning of their random number.\footnote{Of course, a real scheduler
  is unlikely to behave in such an adversarial way. However, 
  if the timing of operations can depend on random values, effects like this
  can arise even with non-adversarial schedulers, as discussed in \Sref{sec:skiplist}.}

In \figref{fig:unbiased} we present a new concurrent version that
\emph{is} statistically unbiased, yet retains the same good properties
of low contention.\footnote{However, it requires $O(\log_2(n))$ bits to
  store the count. Nevertheless, it uses less space than other
  alternatives for decreasing contention (\eg having each thread
  maintain its own local counter).} Our increment function reads the
current value in the counter, then takes the minimum of that value and
a parameter $\MAXCONST$. If the minimum value is $k$, then with
probability $\frac{1}{k+1}$, it uses a fetch-and-add operation
($\fetchAddNoArg$) to atomically add $k+1$ to the counter, otherwise
it returns.
In our version,
the \readfun\ function just returns the value in the counter.  Like
$\compareSwapNoArg$, these $\fetchAddNoArg$ operations are expensive, so
the reason the algorithm scales is that the probability that a
$\fetchAddNoArg$ happens decreases as the counter value grows.
The parameter $\MAXCONST$ caps how small
this probability gets, somewhat like generating only 64
random bits does in the beginning of \figref{fig:dlm}.

How does one show that this algorithm is unbiased, as we have claimed?
Informally, it is because in expectation, each increment adds 1 to the
count, so the total expected value is equal to the number of
increments. Moreover, because addition is associative and commutative,
it does not matter if other threads modify the counter in between when
a $\flipNoArgs$ happens and the corresponding $\fetchAddNoArg$ occurs.
However, it is challenging to make this argument formal. 
We might try to model the value of the counter as a family
of Markov chains,\footnote{Or rather, a \emph{Markov decision process},
  which accounts for the non-determinism of the ordering of
  operations.} as \citeauthor{Flajolet85} did for the
sequential algorithm.  But this is unwieldy because the relevant
state of the chain is not just the current value stored in $l$, but
also the local state of each thread in the middle of an increment
operation.  Moreover, even if one could model the algorithm in this
way, it is hard to justify the connection between the concrete
implementation and this mathematical representation.

As we will see, the program logic we have developed makes this algorithm easy to verify.
\label{subsec:counter}

\subsection{Background from Recent Work}

Our program logic is based on three strands of recent work:

\paragraph{pRHL: probabilistic relational reasoning.}
In many program logics for reasoning about probabilistic programs, assertions
in the logic either explicitly make statements about probabilities, or are interpreted
as being true with some probability (\eg \citep{MorganMS96, BartheGGHS16, KaminskiKMO16, Ramshaw:1979}, among others).
Although effective, this non-standard semantics of assertions is hard to reconcile
with the semantics of concurrent separation logic, in which assertions are understood as claims of ownership of resources.

However, \citeauthor{BartheGHS17} have shown that
reasoning about probabilistic programs can often be done
\emph{without} explicitly reasoning about probabilities in the
assertions of the logic. Using their pRHL logic~\citep{BartheGHS17,
  BartheEGHS17, BartheGB12}, one establishes a refinement
between two randomized programs using proof rules that encode a
special type of simulation relation. The only time that explicit
probabilities arise is a special rule for points in the simulation
when \emph{both} programs take a randomized step. The soundness
theorem for their logic says that derivations in the logic imply the
existence of a \emph{coupling}, a construct from probability theory
that is often used to relate two probability distributions. (We describe couplings in detail in \Sref{sec:couplings}).

\paragraph{Iris: a ``layered'' concurrency logic.}
Modern concurrency logics are rather complex,
making it hard to adapt them
to incorporate probabilistic reasoning. Recent work, however, has
sought to unify and simplify these logics~\citep{views, fcsl, iris}.

In particular, Iris~\citep{iris, iris2, iris3}, a recent higher order
concurrent separation logic, is composed of two layers: a ``base
logic'' and a derived ``program logic'' which is encoded on top of the
base logic. Crucially, most of the difficult semantic constructions are
developed in the base logic.  We are able to encode probabilistic
relational reasoning \emph{\`a la} pRHL in Iris by only modifying the
second layer.  We therefore get all of the results developed in the
base logic ``for free'', and retain the expressive features of Iris.

\paragraph{Indexed valuations: a monadic encoding.}
Polaris, like pRHL, is designed for relational reasoning:
it establishes a refinement between two programs. That means, if we
want to prove a property about some program $e$, we first come up with
some simpler \emph{specification} program $e'$, use the logic to establish
a refinement connecting properties of $e'$ to those of $e$, and then
reason about $e'$.

Therefore, we need to complement our logic with a way to express the
simpler program $e'$ and suitable tools for reasoning about it. We
write these specification programs using the monad of indexed
valuations developed by \citet{VaraccaW06}, which makes it possible to
combine operations for both probabilistic and non-deterministic
choice effects. This monad has a clean equational theory that makes
it possible to reason about probabilistic properties of programs
expressed using it.

\vspace{1em}
Integrating this related work together, however, is not straightforward. The main challenge is that
in pRHL, Hoare ``quadruples'' (which relate a pair of programs, in contrast to a traditional Hoare triple),
are defined quite differently from Hoare triples in Iris. In the former, the quadruple is defined
to hold just when an appropriate coupling between two programs exists. In contrast, in Iris, the Hoare triple is defined by a guarded
fixed point in a step-indexed logic. Roughly speaking, the latter encodes a kind of progress and preservation property:
if the precondition holds, then either the program can take a step (progress), in which case it must do so in a way that leaves
appropriate invariants intact (preservation), or else the program is a value, in which case the postcondition must hold.
Thus, to bring pRHL-style probabilistic relational reasoning to Iris, we had to find a way to translate a condition about whole-program
couplings into a step-by-step property. 

\subsection{Our Contributions}

We make several contributions:

\begin{itemize} 

\item We develop results for reasoning about computations
  expressed in the monadic encoding of \citeauthor{VaraccaW06}. Although prior
  work had used this monad or similar ones combining both effects for denotational semantics~\citep{VaraccaW06, TixKP09a, Goubault-Larrecq07, Goubault-Larrecq15, Mislove06}
  and to reason about small program equivalences~\citep{GibbonsH11},
  we found it necessary to develop new ways to reason about this monad.
  In particular, we develop an (in)equational theory of orderings between
  computations, and rewriting rules for bounding expected values. 
  In addition, we adapt the notion of couplings to this
  setting. Finally, \citeauthor{VaraccaW06} focused on finite probabilistic and
  non-deterministic choice, but their constructions generalize to
  countable probabilistic choice and unbounded non-deterministic choice, which we use.

\item We extend Iris with support for probabilistic relational
  reasoning in the style of pRHL, which lets us establish refinements
  between concurrent programs and these monadic representations.

\item Using Polaris, we prove that the concurrent approximate
  counter algorithm introduced in \Sref{subsec:counter} is
  unbiased.

\item We also verify a fine-grained concurrent two-level skip list,
  and bound the expected number of comparisons performed
  when searching for a key.

\end{itemize}

All of the results in this paper, including the soundness of Polaris
and the examples, have been mechanized in Coq.

We start by describing \citeauthor{VaraccaW06}'s~\citep{VaraccaW06} monad for
probabilistic and non-deterministic choice, and our results for reasoning about
computations expressed in it (\Sref{sec:monad}).  We then describe Polaris
(\Sref{sec:logic}). In \Sref{sec:counters}, we give a detailed explanation of how
we use Polaris to verify the approximate counter example. Then in
\Sref{sec:skiplist} we give an overview of the skip list example.  Finally, we
discuss additional related work in \Sref{sec:related}.

\section{Monadic Representation}
\label{sec:monad}

A common approach to reasoning about effectful programs
is to model effects using a suitable
monad, $M$. Using this monad, one represents an effectful program that
returns a value of type $T$ as a term of type $M(T)$. Next, one
usually proves a series of equational rules for simplifying terms of
type $M(T)$, and other lemmas for reasoning about such
terms. This approach has been used for reasoning about a number of
effects, including: state~\citep{NanevskiMB08, Swierstra09},
non-termination~\citep{CPDT},
non-determinism~\citep{GibbonsH11}, and
probabilistic choice~\citep{AudebaudP09, WeegenM08, PetcherM15, GibbonsH11}.

\subsection{Monads for non-determinism \emph{or} probability.}

Let us recall common monadic encodings
for non-deterministic and probabilistic choice (separately).
For non-determinism, we can define
$\nondetmonad{T}$ as the type consisting of predicates $A: T \rightarrow \PROP$ for which
there exists at least some $t : T$ such that $A(t)$ holds.
We think of these predicates as non-empty sets of terms of type $T$,
where each element of the set represents one of the different non-deterministic outcomes. We say two
terms $A$ and $B$ of type $\nondetmonad{T}$ are equivalent, written $A \equiv B$,
if their sets of elements are the same: for all $x$, $x \in A
\leftrightarrow x \in B$. In addition to the standard monadic
operations (bind and return), we can represent non-deterministic choice between two
computations $A$ and $B$ as the union $A \ndchoice B$ of the two sets, defined by:
\[ A \ndchoice B \equiv \lambda t.\, A(t) \vee B(t) \]
This operation satisfies a number of natural rules:
 \begin{mathpar}
   A \ndchoice B \equiv B \ndchoice A
 
   A \ndchoice (B \ndchoice C) \equiv (A \ndchoice B) \ndchoice C
   
   A \ndchoice A \equiv A
 \end{mathpar}
 These, along with the usual monad laws, can
 be used to prove that one non-deterministic computation is equivalent to another.

We can represent a (discrete) probabilistic computation of type $T$ as a
function $f : T \rightarrow [0, 1]$, mapping values of type $T$ to the
probabilities that they occur. The \emph{support} of $f$, written $\funsupp{f}$
is the set of $t$ such that $f(t) > 0$. Naturally, we want the sum of all the
probabilities for the values in $\funsupp{f}$ to be equal to $1$. In order to
make sense of such an infinite sum, we require the support to be countable. We
define $\probmonad{T}$ to be the type of all functions $f : T \rightarrow [0,
  1]$ such that $\funsupp{f}$ is countable and
\[ \sum_{x \in \funsupp{f}} f(x) = 1\]

Given $A, B: \probmonad{T}$, we say $A \equiv B$ if for all $x$, $A(x) = B(x)$.
We can define an operation which selects between a computation $A$ with probability $p$ and another computation $B$ of the same type with probability $(1-p)$:
\[ A \pchoice{p} B \eqdef \lambda x.\, p\cdot A(x) + (1 - p) \cdot B(x) \]
This operation satisfies equational rules such as:
\begin{mathpar}
  A \pchoice{p} B \equiv B \pchoice{1 - p} A

  A \pchoice{p} A \equiv A
\end{mathpar}

\subsection{Combining effects}
\label{sec:monad-combining}

In order to reason about programs that use both probability and non-determinism,
we would like some way to \emph{combine} the monads we have just described.  We
might try to represent computations of type $T$ combining both effects as terms
of type $\nondetmonad{\probmonad{T}}$, \ie non-empty sets of probability
distributions.

But how do we define the monad operations for this combination? One way to
derive the monad operations for the combination is to specify a \emph{distributive law}~\citep{BeckDistributive}.
However, \citet{VaraccaW06} have given a proof (based on an idea they attribute to Plotkin) that no
distributive law exists between the monads\footnote{More precisely, they consider the case where $\nondetmonadNoArg$ is the monad of \emph{finite} non-empty sets of terms of type $T$, and $\probmonadNoArg$ consists of \emph{finite} distributions, instead of countable ones. However, the impossibility of a distributive law in the finitary case precludes one for the non-finitary versions we have defined.}
we have described above.
For our purposes, it is not necessary to
understand the impossibility proof.
What is important is that, based on their impossibility arguments,
\citeauthor{VaraccaW06} observe that the following equational law, which holds in the probabilistic
choice monad, is problematic if we want to have a distributive law:
\[A \pchoice{p} A \equiv A \]
At first this equivalence seems entirely natural: if in either case we
choose $A$, then the probabilistic choice was irrelevant.
However, when we later add in the effect of non-determinism, removing this law
becomes more justifiable, since it allows us to account for the fact that
subsequent non-determinism in the computation can be \emph{resolved
  differently} on the basis of this seemingly irrelevant probabilistic
choice.

Using this observation, \citeauthor{VaraccaW06} describe an alternative way of
representing probabilistic choice, which they call the \emph{indexed valuation
  monad}, in which this equivalence does not hold, and they then describe a
distributive law between non-empty lists and these indexed valuations to obtain
a monad combining both effects.

An indexed valuation of type $T$ is a tuple $(I, \indfun, \valfun)$,
where $I$ is a countable set whose elements are called \emph{indices}; $\indfun$ is a function
of type $I \to T$, and $\valfun$ is a function of type $I \to
\mathbb{R}$ such that \footnote{In fact,
  \citeauthor{VaraccaW06} first define a more
  general structure in which the sums of $\valfun(i)$ do not have to
  equal 1, and the indices need not be countable. After
  working out some of the theory of these more general objects, they 
  restrict to the subcategory where the indices are finite sets and the probabilities sum to $1$. We will not restrict to finite sets of indices, since by letting them be countable we can model sampling from arbitrary discrete distributions.}
\[ \sum_{i \in I} \valfun(i) = 1 \]
Informally, we can think of the indices as a set of ``codes'' or
identifiers, the $\valfun$ function gives the probability of a
particular index occurring, and $\indfun$ maps these codes to elements of
type $T$. Importantly, the $\indfun$ function is not required to be
injective, so that different codes can lead to the same observable
result. We write $\ivalmonad{T}$ for the type of indexed
valuations of type $T$. The \emph{indicial support} of the valuation,\footnote{\citeauthor{VaraccaW06} call this simply the ``support'' of
  the indexed valuation, however we prefer to use that term for something different, defined below.}
  notated
$\idxsupp{\valfun}$, is the set of indices $i$ for which $\valfun(i) >
0$. We say $(I_1, \indfun_1, \valfun_1) \equiv (I_2,
\indfun_2, \valfun_2)$ if there exists a bijection $h :
\idxsupp{\valfun_1} \rightarrow \idxsupp{\valfun_2}$ such that for all $i
\in \idxsupp{\valfun_1}$, $\valfun_1(i) = \valfun_2(h(i))$ and
$\indfun_1(i) = \indfun_2(h(i))$.  That is, the bijection can only
``relabel'' indices in a way that preserves their probabilities and
what they decode to.
There is a map $\ivaltoprob$ which
takes indexed valuations of type $T$ to elements of $\probmonad{T}$:
\[
  \ivaltoprob(\idxset, \indfun, \valfun)
  = \lambda x.\, \sum_{i \in \indfun^{-1}(\{x\})} \valfun(i)
\]
It is clear that if $\ival_1 \equiv \ival_2$, then
$\ivaltoprob(\ival_1) \equiv \ivaltoprob(\ival_2)$. However, the converse
is not true because $\ival_1$ and $\ival_2$ could have indicial supports with different
cardinalities.

The probabilistic choice between two indexed valuations is defined by:
\[ (\idxset_1, \indfun_1, \valfun_1) \pchoice{p} (\idxset_2, \indfun_2, \valfun_2)
      \eqdef (\idxset_1 + \idxset_2, \indfun', \valfun') \]
where:
\begin{align*}
  \indfun'(i) &=
  \begin{cases}
    \indfun_1(i') & \quad \text{if $i = \inl(i')$} \\
    \indfun_2(i') & \quad \text{if $i = \inr(i')$} 
  \end{cases}
\end{align*}
and
\begin{align*}
  \valfun'(i) &=
  \begin{cases}
    p \cdot \valfun_1(i') & \quad \text{if $i = \inl(i')$} \\
    (1 - p) \cdot \valfun_2(i') & \quad \text{if $i = \inr(i')$}
  \end{cases}
\end{align*}
One can show that for all indexed valuations $\ival_1$ and $\ival_2$
and $0\leq p \leq 1$, we have $\ival_1 \pchoice{p} \ival_2 \equiv
\ival_2 \pchoice{1-p} \ival_1$.

However, unlike the original
probabilistic choice monad we described before, $\ival \pchoice{p}
\ival \nequiv \ival$, unless $p = 0$ or $p = 1$. The reason is that,
when $p$ is neither $0$ nor $1$, the indicial support of $\ival \pchoice{p}
\ival$ will have a larger cardinality than the indicial support of $\ival$, so
there can be no bijection between them. Recall that we do \emph{not} want this equivalence
to hold while trying to define the combined monad, because it is the one that \citet{VaraccaW06} identified as problematic.
With this obstruction removed, it is possible to define
the monad operations on $\nondetmonadNoArg \circ \ivalmonadNoArg$, and we write $\pivalmonadNoArg$ for this composition.
Given $\pival_1$ and $\pival_2$ of type $\pivalmonad{T}$, the probabilistic choice
operation $\pival_1 \pchoice{p} \pival_2$ is defined by taking the pairwise probabilistic choice of each indexed valuation in the respective sets: 
\[ \pival_1 \pchoice{p} \pival_2 \equiv \{ \ival_1 \pchoice{p} \ival_2 \ | \ \ival_1 \in \pival_1, \ival_2 \in \pival_2 \}
\]
while the non-deterministic choice is once again the union of the sets.
We say $\pival_1 \equiv \pival_2$ if for each $\ival_1 \in \pival_1$, there exists some $\ival_2 \in \pival_2$ such that $\ival_1 \equiv \ival_2$, and vice versa.
The full definition of the bind operation is somewhat involved, as are the proofs of the monad laws, so we refer to \citet{VaraccaW06}. What is important is the equational properties that hold, of which a selection are shown in \figref{fig:monad-eqn-laws} (the standard monad laws are omitted).

\subsection{Example: Modeling approximate counters}
\newcommand\monadeqnsfig{\begin{figure}
  \begin{mathpar}
    \pival_1 \pchoice{p} \pival_2 \equiv \pival_2 \pchoice{1-p} \pival_1
    \and
    \pival_1 \pchoice{1} \pival_2 \equiv \pival_1
    \and
    \pival \ndchoice \pival \equiv \pival
    \and
    \pival_1 \ndchoice \pival_2 \equiv \pival_2 \ndchoice \pival_1
    \and
    \pival_1 \ndchoice (\pival_2 \ndchoice \pival_3)
    \equiv (\pival_1 \ndchoice \pival_2) \ndchoice \pival_3
    \and
    \pival_1 \pchoice{p} (\pival_2 \ndchoice \pival_3)
      \equiv (\pival_1 \pchoice{p} \pival_2) \ndchoice (\pival_1 \pchoice{p} \pival_3)
    \and
    \mbind{x}{\pival_1 \ndchoice \pival_2}{F(x)}
    \equiv \left(\mbind{x}{\pival_1}{F(x)}\right) \ndchoice
           \left(\mbind{x}{\pival_2}{F(x)}\right)
    \and
    \mbind{x}{\pival_1 \pchoice{p} \pival_2}{F(x)}
    \equiv \left(\mbind{x}{\pival_1}{F(x)}\right) \pchoice{p}
           \left(\mbind{x}{\pival_2}{F(x)}\right)
  \end{mathpar}
\caption{Equational laws for $\nondetmonadNoArg \circ \ivalmonadNoArg$ monad.}
\label{fig:monad-eqn-laws}
\end{figure}
}

\newcommand\extremafig{\begin{figure}
    \begin{mathpar}
    \infer{}{\exMin{f}{\mret{v}} = f(v)}
    \and
      \infer{k \geq 0}{\exMin{(\lambda x. k \cdot f(x) + c)}{\pival} = k \cdot \exMin{f}{\pival} + c}
    \and
    \infer{}{\exMin{f}{\pival_1 \pchoice{p} \pival_2} =
      p\cdot \exMin{f}{\pival_1} + (1 - p) \cdot \exMin{f}{\pival_2}}
    \and
    \infer{}{\exMin{g \circ f}{\pival} = \exMin{g}{\mbind{x}{\pival}{\mret{f(x)}}}}
    \and
    \infer{\forall x.\, k_1 \leq \exMin{f}{F(x)} \leq k_2}{k_1 \leq \exMin{f}{\mbind{x}{\pival_1}{F(x)}} \leq k_2}
    \end{mathpar}
    \caption{Selection of rules for calculating extrema of expected values (analogous rules for $\exMax{f}{-}$ omitted).}
    \label{fig:extrema}
\end{figure}
}

\newcommand\monadorderingfig{\begin{figure}
  \begin{mathpar}
    \infer{\pival_1 \equiv \pival_2}{\pival_1 \pivalleq \pival_2}
    \and
    \infer{\pival_1 \pivalleq \pival_2 \\ \pival_2 \pivalleq \pival_1}{\pival_1 \equiv \pival_2}
    \and
    \infer{\pival_1 \pivalleq \pival_2 \\ \pival_2 \pivalleq \pival_3}
          {\pival_1 \pivalleq \pival_3}
    \and
    \infer{\pival_1 \pivalleq \pival_1' \\
           \pival_2 \pivalleq \pival_2'}
          {\pival_1 \pchoice{p} \pival_2 \pivalleq \pival_1' \pchoice{p} \pival_2'}
    \and
    \infer{\pival_1 \pivalleq \pival_1' \\
           \pival_2 \pivalleq \pival_2'}
          {\pival_1 \ndchoice \pival_2 \pivalleq \pival_1' \ndchoice \pival_2'}
    \and
    \infer{}{\pival_1 \pivalleq \pival_1 \ndchoice \pival_2}
    \and
    \infer{\pival_1 \pivalleq \pival_2 \\
           \forall x.\, F_1(x) \pivalleq F_2(x)}
          {\mbind{x}{\pival_1}{F_1(x)} \pivalleq \mbind{x}{\pival_2}{F_2(x)}}
    \and
  \end{mathpar}
\caption{Rules for ordering on $\nondetmonadNoArg \circ \ivalmonadNoArg$.}
\label{fig:monad-ordering-laws}
\end{figure}
}

\newcommand\monadirrelfig{\begin{figure}
  \begin{mathpar}
    \infer{}{\ival \irrelequiv \ival}
    \and
    \infer{\ival_1 \equiv \ival_1' \\
      \ival_2 \equiv \ival_2' \\
      \ival_1 \irrelequiv \ival_2}
      {\ival_1' \irrelequiv \ival_2'}
    \and
    \infer{\ival_1 \irrelequiv \ival_2 \\
           \ival_2 \irrelequiv \ival_3}
          {\ival_1 \irrelequiv \ival_3}
    \and
    \infer{\ival_1 \irrelequiv \ival_2 \\
           \forall x.\, F_1(x) \irrelequiv F_2(x)}
          {\mbind{x}{\ival_1}{F_1(x)} \irrelequiv \mbind{x}{\ival_2}{F_2(x)}}
    \and
    \infer{\ival_1 \irrelequiv \ival_2 \\ \ival_1' \irrelequiv \ival_2'}
          { \ival_1 \pchoice{p} \ival_1' \irrelequiv \ival_2 \pchoice{p} \ival_2'}
    \and
    \infer{}
          {(\mbind{x}{\ival_1}{\ival_2}) \irrelequiv \ival_2}
  \end{mathpar}
\caption{Rules for the $\irrelequiv$ relation on indexed valuations.}
\label{fig:irrel-equiv}
\end{figure}
}

\newcommand\monadirrelleqfig{\begin{figure}
  \begin{mathpar}
    \infer{}{\pival \irrelleq \pival}
    \and
    \infer{\pival_1 \irrelleq \pival_2 \\
           \pival_2 \irrelleq \pival_3}
           {\pival1 \irrelleq \pival_3}
    \and
    \infer{\pival_1' \pivalleq \pival_1 \\
      \pival_2 \pivalleq \pival_2' \\
      \pival_1 \irrelleq \pival_2}
      {\pival_1' \irrelleq \pival_2'}
    \and
    \infer{\pival_1 \irrelleq \pival_2 \\
           \forall x.\, F_1(x) \irrelleq F_2(x)}
          {\mbind{x}{\pival_1}{F_1(x)} \irrelleq \mbind{x}{\pival_2}{F_2(x)}}
    \and
    \infer{\pival_1 \irrelleq \pival_2 \\ \pival_1' \irrelleq \pival_2'}
          { \pival_1 \pchoice{p} \pival_1' \irrelequiv \pival_2 \pchoice{p} \pival_2'}
    \and
    \infer{}
          {(\mbind{x}{\pival_1}{\pival_2}) \irrelleq \pival_2}
  \end{mathpar}
\caption{Rules for $\irrelleq$ relation.}
\label{fig:irrel-leq}
\end{figure}
}

\newcommand\couplingrulesfig{\begin{figure}
  \begin{mathpar}
    \inferH{Ret}{P(a, b)}{\ipCoupling{\mret{a}}{\mret{b}}{P}}
    \and
    \inferH{Equiv}{\ival \equiv \ival' \\
           \pival \pivalleq \pival' \\
           \ipCoupling{\ival}{\pival}{P}}
          {\ipCoupling{\ival'}{\pival'}{P}}
     \and
     \inferH{Conseq}{\ipCoupling{\ival}{\pival}{P} \\
       \forall x, y.\, P(x, y) \rightarrow P'(x, y)}
          {\ipCoupling{\ival}{\pival}{P'}}
     \and
    \inferH{Bind}{\ipCoupling{\ival}{\pival}{P} \\
           \forall x, y.\, P(x, y) \rightarrow \ipCoupling{F(x)}{F'(y)}{Q}}
          {\ipCoupling{(\mbind{x}{\ival}{F(x)})}{(\mbind{y}{\pival}{F'(y)})}{Q}}
    \and
    \inferH{P-Choice}{\ipCoupling{\ival}{\pival}{P} \\
           \ipCoupling{\ival'}{\pival'}{P}}
          {\ipCoupling{\ival \pchoice{p} \ival'}{\pival \pchoice{p} \pival'}{P}}
    \and
    \inferH{Trivial}{}
          {\ipCoupling{\ival}{\pival}{\TRUE}}
    \and
  \end{mathpar}
\caption{Rules for constructing non-deterministic couplings.}
\label{fig:coupling-rules}
\end{figure}
}

 \monadeqnsfig
\newcommand\monadcoqfig{\begin{figure}
\lstinputlisting{monad-coq.v}
\caption{Monadic encoding of approximate counter algorithm from \figref{fig:unbiased}.}
\label{fig:monad-coq}
\end{figure}
}

\newcommand{\approxN}{\textlog{approxN}}
\newcommand{\approxIncr}{\textlog{approxIncr}}

\newcommand\monadcounterfig{\begin{figure}
    \begin{minipage}[t]{0.4\linewidth}
    \[
    \begin{array}{ll}
      &\approxIncr \eqdef\\
      &\quad \mbind{k}{\mret{0} \ndchoice \cdots \\
         & \quad\quad\quad \ndchoice\ \mret{\MAXCONST}}{} \\
      &\quad \mret{(k+1)} \pchoice{\frac{1}{k+1}} \mret{0}
    \end{array}
    \]
    \end{minipage}%
    \qquad
    \vrule
    \qquad
    \begin{minipage}[t]{0.4\linewidth}
    \[
    \begin{array}{ll}
      &\approxN\ 0\ l \eqdef \mret{l} \\
      &\approxN\ (n + 1)\ l \eqdef \\
      &\quad \mbind{k}{\approxIncr}{} \\
      &\quad \approxN\ n\ (l + k)
    \end{array}
    \]
    \end{minipage}

\caption{Monadic encoding of approximate counter algorithm from \figref{fig:unbiased}.}
\label{fig:monad-counter}
\end{figure}
}
 
In \figref{fig:monad-counter} we show how to model the approximate
counter code from \figref{fig:unbiased} using this monad.  The
$\approxIncr$ computation first non-deterministically selects a number
{\tt k} up to $\MAXCONST$ -- this models the process of taking the
minimum of the value in $l$ and $\MAXCONST$ in the code. The
non-determinism accounts for the fact that the value that will be read
depends on what other threads do. The monadic encoding then
makes a probabilistic choice, returning $k+1$ with probability
$\frac{1}{k+1}$ and $0$ otherwise, which represents the
probabilistic choice that the code will make about
whether to do the fetch-and-add.

Finally, the process of repeatedly incrementing the counter $n$
times is modeled by $\approxN$. The first argument $n$ tracks
the number of pending increments to perform, while the second argument
$l$ accumulates the sum of the values returned by the calls to
$\approxIncr$. Note that this model \emph{does not} try to represent
multiple threads in the middle of an increment each
waiting to add its value to the shared count -- rather, it is \emph{as if}
the actual calls to {\tt incr} all happened atomically in sequential order,
with the effects of concurrency captured by the non-determinism in the $\approxIncr$
computation.

Of course, we need to show that this model accurately captures the behavior of the code from
\figref{fig:unbiased} -- this is what the program logic we describe in \Sref{sec:logic}
will do. First, however, we need to describe the new results we have developed to reason further about
the monadic encoding itself.

\monadcounterfig
\subsection{Reasoning about Quantitative Properties}
\label{sec:monad-quant}

With what we have described so far, we can express
computations with randomness and non-determinism and derive equivalences between them, but
we do not yet have a way to talk about the standard concerns of probability theory (\eg expected values, variances, tail bounds).

Given an indexed valuation $\ival = (\idxset, \indfun, \valfun)$ of type $T$
and a function $f : T \to \mathbb{R}$, we can define the
\emph{expected value} of $f$ on $\ival$ as:
\[ \exival{f}{\ival} \eqdef \sum_{i \in \idxset} f(\indfun(i)) \cdot \valfun(i) \]
(this coincides with the usual notion of expected value of a
random variable if we interpret the indexed valuation as a
distribution using the map $\ivaltoprob$ defined above).
Since $\idxset$ is a countable set, the above
series may not necessarily converge. We say that the expected value
of $f$ on $\ival$ exists if the above series converges absolutely.
Throughout this paper, when we mention expected values in rules and derivations,
we will implicitly assume side conditions stating that all the relevant expected values exist.

If $\ival_1 \equiv \ival_2$, then for all $f$, $\exival{f}{\ival_1} = \exival{f}{\ival_2}$.
Moreover, given a value $t$ of type $T$, if we define $[t]$ to be the
\emph{indicator function} that returns $1$ if its input is equal to
$t$ and $0$ otherwise, then $\exival{[t]}{\ival}$ is equal to the
probability that $\ival$ yields the value $t$, so we can encode probabilities
as expected values.

Since an $\pival$ of type $\pivalmonad{T}$ is just a set
of indexed valuations, we can apply $\exival{f}{-}$ to each $\ival
\in \pival$ to get the set of expected values that can arise
depending on how non-deterministic choices are resolved. Generally
speaking, we will be interested in bounding the smallest or largest
possible value that these expected values can take. We can define the
\emph{minimal} and \emph{maximal} expected value of $f$ on $\pival$ as:
\begin{mathpar}
\exMin{f}{\pival} \eqdef \inf_{\ival \in \pival} \exival{f}{\ival} \and
\exMax{f}{\pival} \eqdef \sup_{\ival \in \pival} \exival{f}{\ival}
\end{mathpar}
We say that these extrema exist if for all $\ival \in \pival$,
$\exival{f}{\ival}$ exists. Since $\pival$ may be an infinite set,
$\exMin{f}{\pival}$ and $\exMax{f}{\pival}$ can be $-\infty$ and $+\infty$
respectively.  Rules for calculating these values are given in
\figref{fig:extrema}. As before, we implicitly assume that all of the stated
extrema exist and are finite.

\extremafig

To help reason about these extrema, we introduce a
partial order on terms of type $\pivalmonad{T}$: We
say $\pival_1 \pivalleq \pival_2$ if for each $\ival_1 \in \pival_1$, there
exists some $\ival_2 \in \pival_2$ such that $\ival_1 \equiv
\ival_2$. If $\pival_1 \pivalleq
\pival_2$ then $\exMax{f}{\pival_1} \leq \exMax{f}{\pival_2}$ and
$\exMin{f}{\pival_2} \leq \exMin{f}{\pival_1}$. Thus, we can bound
$\pival_1$'s extrema by first finding some $\pival_2$ such that
$\pival_1 \pivalleq \pival_2$, and then bounding the latter's extrema.

One way to ensure that expected values exist 
is to show that the functions we are computing expected values of are
suitably bounded.
We first define the \emph{support} of $\ival$ as the set of all values
that occur with non-zero probability:
  \[ \Isupp{\ival} \eqdef \{ v \ | \ \exists i \in \ival.\, \indfun(i) = v \wedge \valfun(i) > 0 \} \]
The support of a set of indexed valuations $\pival$ is then the union of their
supports:
  \[ \Psupp{\pival} \eqdef \bigcup\limits_{\ival \in \pival} \Isupp{\ival} \]
We say that $f$ is bounded on the support of $\pival$ if there exists some $c$
such that $|f(v)| \leq c$ for all $v \in \Psupp{\pival}$. If this holds,
then $\exMin{f}{\pival}$ and $\exMax{f}{\pival}$ exist and are finite.

\vspace{2em}
Using the above rules, we can show that $\exMin{\textlog{id}}{\approxN\ n\
  0} = \exMax{\textlog{id}}{\approxN\ n\ 0} = n$, which implies that no
matter how the non-determinism is resolved in our model of the
counter, the expected value of the result will be the
number of increments.  Let us just consider the case for the minimum,
since the maximum is the same. The proof proceeds by induction on $n$,
after first strengthening the induction hypothesis to the claim that
$\exMin{\textlog{id}}{\approxN\ n\ l} = n + l$. The key step of
the proof is to show that $\exMin{\textlog{id}}{\approxIncr} =
1$, \ie each increment contributes $1$ to the expected value. From the
last rule in \figref{fig:extrema}, it suffices to show that whatever
value of $k$ is non-deterministically selected, the resulting
expected value will be $1$. We have that for
all $k$:
\begin{align*}
  &\exMinBig{\textlog{id}}{\mret{(k+1)} \pchoice{\frac{1}{k+1}} \mret{0}} \\
  &= \left(\frac{1}{k+1}\right) \cdot (k+1) + \left(1 - \frac{1}{k+1}\right) \cdot 0 \\
  &= 1
\end{align*}

Let us summarize the discussion so far. Because the non-determinism monad
$\nondetmonadNoArg$ could not be combined with the standard probabilistic choice
monad $\probmonadNoArg$, we replaced the latter with the monad of indexed
valuations, $\ivalmonadNoArg$. The distinction between $\ivalmonadNoArg$ and
$\probmonadNoArg$ is that indexed valuations carry additional data (the indices)
and have a finer notion of equivalence. Because of this finer equivalence, the problematic
equational law does not hold for $\ivalmonadNoArg$, so that the monad operations can
be defined on $\pivalmonadNoArg$.

We then re-developed the notions of expected values and probabilities for
$\ivalmonadNoArg$ and defined corresponding extrema of expected values for
$\pivalmonadNoArg$. Since these definitions respect the equivalence relations on
$\ivalmonadNoArg$ and $\pivalmonadNoArg$, we could bound the extrema of some
$\pival$ by first finding $\pival'$ such that $\pival \equiv \pival'$,
and then bounding the extrema of $\pival'$. More generally, we also had the $\pivalleq$
relation, so that similar bounds could be obtained solely by showing $\pival \pivalleq \pival'$.

However, the relations $\equiv$ and $\pivalleq$ above are finer than necessary
if our goal is to use them to translate bounds on extrema of $\pival'$ to bounds
on $\pival$, and similarly so for translating bounds on expected values of one indexed valuation to another. For example, if $f$ is a bounded function, then
$\exival{f}{\ival} = \exival{f}{\ival \pchoice{p} \ival}$,
yet we know that $\ival \not\equiv \ival \pchoice{p} \ival$.

Because we will be interested in using relational reasoning in order to bound expected values,
it is useful to define the coarsest relations that suffice for this purpose, and
derive some properties about them. We define $\ival \irrelequiv \ival'$ to hold
if for all bounded\footnote{The reason for quantifying over bounded functions is
  to ensure that the two expected values exist.} functions $f$,
$\exival{f}{\ival} = \exival{f}{\ival'}$. Because indicator functions are bounded,
observe that if $\ival \irrelequiv \ival'$, 
then $\exival{[t]}{\ival} = \exival{[t]}{\ival'}$ for all $t$. In other words, this notion of
equivalence is the same as saying that the probability distributions $H(\ival)$
and $H(\ival')$ are equal.
Rules for this relation are shown in \figref{fig:irrel-equiv}.
Crucially, it is a congruence with respect to all the monad operations.
 
 \monadirrelfig

Analogously, we define $\pival \irrelleq \pival'$ to hold if for all bounded
functions $f$, $\exMax{f}{\pival} \leq \exMax{f}{\pival'}$. Thus, if this
relation holds, one can bound the maxima of $\pival$ by bounding the maxima of
$\pival'$. Since the negation of $f$ is bounded if and only if $f$ is, this also
implies that $\exMin{f}{\pival'} \leq \exMin{f}{\pival}$ for all bounded
$f$. Some rules for this relation are shown in \figref{fig:irrel-leq}.
 
 \monadirrelleqfig
 
 The following lemma shows that the definition of $\irrelleq$ generalizes to a larger
 class of functions than just the bounded ones:

 \begin{lemma}
   If $\pival \irrelleq \pival'$ and $f$ is bounded on the support of $\pival'$,
   then $f$ is bounded on the support of $\pival$ and
  $\exMax{f}{\pival} \leq \exMax{f}{\pival'}$.
 \end{lemma}

\subsection{Nondeterministic Couplings}
\label{sec:couplings}

Recent work by
\citeauthor{BartheEGHSS15}~\citep{BartheEGHSS15, BartheGHS17,
  BartheEGHS17} has shown that the notion of a probabilistic \emph{coupling}~\citep{lindvall2002lectures} is fundamental for
relational reasoning in probabilistic program logics. Given
two distributions $A : \probmonad{T_A}$ and $B : \probmonad{T_B}$,
a coupling between $A$ and $B$ is a distribution $C : \probmonad{T_A \times T_B}$
such that:
\begin{enumerate}
\item $\forall x : T_A.\, A(x) =  \sum_{y} C(x, y)$
\item $\forall y : T_B.\, B(y) =  \sum_{x} C(x, y)$
\end{enumerate}
That is, $C$ is a joint distribution whose marginals equal $A$ and $B$.
These two conditions are equivalent to requiring that:
\begin{enumerate}
\item  $A \equiv \left(\mbind{(x, y)}{C}{\mret{x}}\right)$
\item  $B \equiv \left(\mbind{(x, y)}{C}{\mret{y}}\right)$
\end{enumerate}
Given a predicate $P : A \times B \rightarrow \Prop$, we
say that $C$ is a $P$-coupling, if, in addition to the above, we have:
\[ \forall x, y.\, C(x, y) > 0 \rightarrow P(x, y) \]
\ie all pairs $(x, y)$ in the support of the distribution $C$ satisfy $P$. 
The existence of a $P$-coupling can tell us important things about the
two distributions. For example, if $P(x, y) = (x = y)$, then the
existence of a $P$-coupling tells us the two distributions are
equivalent. Moreover, there are rules for systematically constructing
couplings between distributions. We will explain some of these rules once
we have described how to adapt couplings to the monad $\pivalmonadNoArg$.

First, using the monadic formulation of the coupling conditions, it is
straightforward to define an analogous idea for $\ivalmonadNoArg$:
Given $\ival_1 : \ivalmonad{T_1}$ and $\ival_2 : \ivalmonad{T_2}$, a
coupling between $\ival_1$ and $\ival_2$ is an $\ival : \ivalmonad{T_1
  \times T_2}$ such that:
\begin{enumerate}
\item  $\ival_1 \irrelequiv \left(\mbind{(x, y)}{\ival}{\mret{x}}\right)$
\item  $\ival_2 \irrelequiv \left(\mbind{(x, y)}{\ival}{\mret{y}}\right)$
\end{enumerate}
and $\ival = (\idxset, \indfun, \valfun)$ is a $P$-coupling if for all
$i$ such that $\valfun(i) > 0$, $P(\indfun(i))$ holds. As before, if
$P$ is the equality predicate, then the existence of a $P$-coupling
between $\ival_1$ and $\ival_2$ implies $\ival_1 \irrelequiv \ival_2$. 

We can lift this to a relation between a single indexed valuation
$\ival$ and a set of indexed valuations $\pival$. We
say\footnote{\citet{BartheEGHSS15} use ``non-deterministic
  coupling'' to refer to a particular kind of
  coupling which is unrelated to adversarial non-deterministic choice.} there is a
\emph{non-deterministic $P$-coupling} between $\ival$ and $\pival$ if
there exists some $\ival'$ such that $\{ \ival' \} \irrelleq \pival$ and a $P$-coupling between
$\ival$ and $\ival'$. We write $\ipCoupling{\ival}{\pival}{P}$ to
denote the existence of such a coupling.

Rules for constructing these couplings are shown in \figref{fig:coupling-rules}.
If we interpret the $P$ in $\ipCoupling{\ival}{\pival}{P}$ 
as a kind of ``post-condition'' for the execution of the computations $\ival$ and $\pival$,
then these coupling rules have the structure of a Hoare-like relational logic~\citep{benton:popl04},
as in the work of \citet{BartheEGHSS15}: \eg the rule \ruleref{Bind} is analogous to the usual sequencing rule in Hoare logic.

The rule \ruleref{P-Choice} lets us couple probabilistic choices
$\ival \pchoice{p} \ival'$ and $\pival \pchoice{p} \pival'$ with
post-condition $P$ by coupling $\ival$ to $\pival$ and $\ival'$ to
$\pival'$. This is somewhat surprising: we get to reason about
these two probabilistic choices as if they both chose the left
alternative or both chose the right alternative, rather than considering
the full set of four combinations. This counter-intuitive rule is quite
useful, as demonstrated in the many examples given in the work of \citeauthor{BartheEGHSS15}
We will see an example of its use in \Sref{sec:counters}.

\couplingrulesfig

The following theorem lets us use the existence of a non-deterministic coupling to bound expected values:
 \begin{theorem}
  \label{thm:irrel-coupling-bound}
   Let $g$ be bounded on $\Psupp{\pival}$ and let $P(x, y) = (f(x) = g(y))$.
   If\ $\ipCoupling{\ival}{\pival}{P}$, then $\exival{f}{\ival}$ exists and
   \[ \exMin{g}{\pival} \leq \exival{f}{\ival} \leq \exMax{g}{\pival} \]
   \label{thm:coupling}
 \end{theorem}
\section{Program Logic}
\label{sec:logic}
We now describe Polaris, the program logic we have developed for proving that a
program is modeled by the monadic specifications from the previous
section.

\subsection{Program Semantics}

Polaris is parameterized by a generic probabilistic concurrent
language. However, for concreteness, we instantiate it with the
ML-like language used in the examples from \Sref{sec:intro}.
\figref{fig:lang} gives the syntax and semantics of this language. We
omit the standard rules for things like tuples, recursive functions,
and references.  The per-thread reduction relation $\expr; \state
\pstep{p} \expr'; \state'$ is annotated with a probability $p$ that
the transition takes place. We say $\expr$ is atomic, written $\atomic(\expr)$,
if $\expr$ reduces to a value in a single step.

\begin{figure}
\noindent \textbf{Syntax:} 
 \begin{mathpar}
 \begin{array}{llcl}
 \textdom{Val} & \val & ::= & 
    \lambda \var.\, \expr_1 \ALT
    (\val_1, \val_2) \ALT 
    \vunit \ALT 
    n \ALT  
    b \ALT \dots  \\
 \textdom{Expr} & \expr & ::= & \var \ALT 
    \val \ALT
    \expr_1 \, \expr_2 \ALT
    \fork{\expr} \ALT
    \flip{\expr_1, \expr_2} \ALT
    \dots \\
 \textdom{Eval Ctx} & \lctx & ::= & [] \ALT
    \lctx \, \expr \ALT 
    \Val \, \lctx \ALT 
    \flip{\lctx, \expr} \ALT 
    \flip{\val, \lctx} \ALT 
    \dots \\
 \textdom{State} & \state & \in & \mathbb N \rightarrow \textdom{Val}  \\
 \textdom{Config} & \cfgvar & \in &\{ l : \textdom{List Expr} \ | \ l \neq \emptyset \}
   \times \textdom{State}  \\
 \textdom{Trace} & \trace & \in & \{ l : \textdom{List Config} \ | \ l \neq \emptyset \} \\
 \textdom{Scheduler} & \sched & \in & \textdom{Trace} \rightarrow \mathbb{N}
 \end{array}
 \end{mathpar}

\noindent \textbf{Per-Thread Reduction:} $\expr; \state \pstep{p} \expr'; \state'$

\begin{mathpar}
  \inferH{Flip-True}{0 \leq \frac{n_1}{n_2} \leq 1}
         {\flip{n_1, n_2}; \state \pstep{\frac{n_1}{n_2}} \TRUE; \state}

  \inferH{Flip-False}{0 \leq \frac{n_1}{n_2} \leq 1}
         {\flip{n_1, n_2}; \state \pstep{1 - \frac{n_1}{n_2}} \FALSE; \state}

\text{(Standard rules omitted.)}
\end{mathpar}

\noindent \textbf{Concurrent Semantics:} $\cfgvar \pistep{p}{i} \cfgvar'$
\begin{mathpar}
\infer{\expr_i ; \state \pstep{p} \expr_i' ; \state' }
      {\cfg{\dots,\lctx[\expr_i],\dots}{\state}
       \pistep{p}{i}
       \cfg{\dots,\lctx[\expr_i'],\dots}{\state'}
      }

\infer{}
      {\cfg{\expr_1,\dots,\expr_{i-1},\lctx[\fork{\expr_\f}],\dots}{\state}
       \pistep{1}{i}
       \cfg{\expr_1,\dots,\expr_{i-1},\lctx[\vunit],\dots,\expr_\f}{\state}
      }
\end{mathpar}

\noindent \textbf{Trace Semantics}: $\trace \trstep{p}{\sched} \trace'$

\begin{mathpar}
\infer{\sched(\trace,\cfgvar) = i \\
       \cfgvar \pistep{p}{i} \cfgvar'}
      {\trace,\cfgvar \trstep{p}{\sched} \trace,\cfgvar,\cfgvar'}

\infer{\sched(\trace,\cfgvar) = i \\
       \neg(\exists \cfgvar', p.\, \cfgvar \pistep{p}{i} \cfgvar')}
      {\trace,\cfgvar \trstep{1}{\sched} \trace,\cfgvar,\cfgvar}

\end{mathpar}

\caption{Syntax and semantics of concurrent language.}
\label{fig:lang}
 \end{figure}
 
The $\flip{n_1, n_2}$ command takes two integers as arguments and
simulates a biased coin flip: it transitions to $\TRUE$ with
probability $\frac{n_1}{n_2}$ and $\FALSE$ with probability $1 -
\frac{n_1}{n_2}$. (In the introduction we somewhat informally wrote
$\flip{n_1 / n_2}$ as if the language had rational numbers as a
primitive). There is a side condition to ensure that $\frac{n_1}{n_2}$
actually corresponds to a valid probability. Other than this command,
the per-thread transition system for this language is
deterministic.
The generic framework for our logic allows us to extend
this language with other probabilistic commands, so long as they only
sample from discrete probability distributions.

This per-thread reduction relation is then lifted to a concurrent
transition system.  A \emph{configuration} $\cfgvar$ is a pair
consisting of a list of expressions (representing a pool of threads)
and a state $\state$.  We say $\cfgvar \pistep{p}{i} \cfgvar'$
when the $i^{th}$ thread of $\cfgvar$ transitions with probability $p$
leading to a new configuration $\cfgvar'$. The
$\fork{\expr_\f}$ command adds a new thread $\expr_\f$ to the
pool.

A \emph{scheduler} decides which thread will get to step at each point
in an execution. We model a scheduler as a function $\varphi$ of type
$ \textdom{Trace} \rightarrow \mathbb{N}$, where a
trace is a non-empty list of configurations representing a partial
execution. The scheduler is permitted to inspect the entire
history and complete state of the program when deciding which thread
gets to go next. Of course, a real implementation of a scheduler does
not actually do this, but by conservatively considering this strong
class of \emph{adversarial} schedulers, results we prove will also
hold for realistic schedulers. We write $\trace \trstep{p}{\sched}
\trace'$ to indicate that the thread selected by $\sched(\trace)$
steps with probability $p$ to a new configuration which is appended to
$\trace$ to obtain $\trace'$. We write $\curr(\trace)$ for the last configuration in a trace.
If the scheduler returns a
thread number which cannot take a step or which does not exist, the system takes a ``stutter''
step and the current configuration is repeated again at the end of the
trace.
We say $\trace$ reduces to $\trace'$ in $n$ steps under $\sched$ if:
\[ \trace \trstep{p_1}{\sched} \dots \trstep{p_{n}}{\sched} \trace' \]
for some $p_1, \dots, p_n$ where each $p_i > 0$.
A configuration $\cfgvar$ has terminated if the first thread in the pool is a value.
We say that $\trace$ is terminating in at most $n$ steps under $\sched$, if for all $\trace'$
which $\trace$ reduces to under $\sched$ for $n' \geq n$ steps, $\curr(\trace')$ has terminated.

We now want to interpret this reduction relation as defining a distribution on
program executions. However, in general, this would require measure theoretic probability to handle properly: even though our language
only features sampling from countable discrete distributions, the set of all
executions of a program is uncountable if the program does not necessarily
terminate.\footnote{A more denotational alternative, based on an approach due to \citet{Kozen81}, is to
  interpret programs as monotone maps on sub-distributions of states. Then recursive commands are interpreted as least fixed points. However, since the original soundness proof of Iris is given in terms of a language with an operational semantics, we found it easier to use the semantics we describe in this section.}
However, if we restrict consideration to programs that terminate in a bounded
number of steps, we can avoid these issues. Since most concurrency logics only
handle partial correctness specifications anyway, this does not lead to much further loss of generality.

With this restriction in place, we can interpret program executions as indexed valuations (as we explained in \Sref{sec:monad}, indexed valuations can be interpreted as probability distributions, and vice versa). Given a scheduler
$\sched$ and a trace $\trace$, we first convert the trace step relation to an
indexed valuation. Since the set of traces $\trace'$ which $\trace$ can step to
under $\sched$ is countable, we can take the set of indices $\idxset$ to be
any set in bijection with this set of traces. Take $\indfun$ to be this
bijection, and set $\valfun(i)$ equal to the probability $p$ such that $\trace
\trstep{p}{\sched} \indfun(i)$.  We refer to the resulting indexed valuation
$(\idxset, \indfun, \valfun)$ as $\trstepival{\sched}{\trace}$. For each $n$, we
define the indexed valuation $\trstepivalN{\sched}{\trace}{n}$ recursively by:
\[
\begin{array}{lll}
  \trstepivalN{\sched}{\trace}{0} & \eqdef &
  \match{\curr(\trace)}\ ([\expr_1, \dots], \state) => \mret{\expr_1} \,
  \matchend \\
   \trstepivalN{\sched}{\trace}{n + 1} & \eqdef& \mbind{\trace'}{\trstepival{\sched}{\trace}}{\trstepivalN{\sched}{\trace'}{n}}
\end{array}
\]
This corresponds to stepping the trace $n$ times and returning the
first thread from the final configuration of the resulting trace. We
regard the ``return value'' of a concurrent program to be the value
that the first thread evaluates to, so in the event that the program
terminates in $n$ steps under the scheduler,
$\trstepivalN{\sched}{\trace}{n}$ gives this return value.

\subsection{Background on Iris}

As we have mentioned, Polaris is an extension of Iris, a
recent concurrency logic with many expressive features. For reasons of
space, we cannot explain all of Iris. We
refer the reader to the Iris papers and manual \citep{iris, iris2,
  iris3, IrisManual} for a full account.  Instead, we will just describe some
essential aspects needed to understand our extensions and
examples.

\figref{fig:hoarerules} shows the basic concurrent separation logic
rules of Iris. (Treat the $\vs$ connective as just a kind of
implication for now, we explain its use below.)
These rules are used to establish triples of the form:
\[ \hoare{\prop}{\expr}{x.\, \propB} \]
which imply that if $\expr$ is executed in a state that initially
satisfies $\prop$, and it terminates with value $\val$, then the
terminating state will satisfy $[\val/x]\propB$. Furthermore, at no point will
$\expr$ go wrong and reach a stuck state, and neither will any of the
threads forked by $\expr$ during its execution. This is a partial
correctness property: the post-condition only holds under executions where $\expr$
terminates.

The fundamental idea of separation logic~\citep{reynolds:separation} is the
separating conjunction $P * Q$, which says
that the program heap can be split into two disjoint pieces satisfying
$P$ and $Q$ respectively. Thus, assertions are interpreted as claims
of \emph{ownership of resources}, where a ``resource'' is just a
fragment of the heap. In particular, the $l \mapsto v$ assertion
claims ownership of a part of the heap containing the location $l$,
and moreover says that $l$ maps to the value $v$. Ownership of this
resource licenses a thread to access and modify this location (see
\ruleref{ml-load} and \ruleref{ml-store} in \figref{fig:hoarerules}).

\newcommand\irisrulesfig{\begin{figure}
\begin{mathpar}
\inferH{ml-alloc}{}
       {\hoare{\TRUE}{\alloc{\val}}{\Ret\var. \var\mapsto\val}[]}

\inferH{ml-load}{}
{\hoare{\loc \mapsto \val}{\load\loc}{\Ret\var. \var = \val \wedge \loc \mapsto \val}[]}

\inferH{ml-store}{}
{\hoare{\loc \mapsto \val}{\store\loc\valB}{\loc \mapsto \valB}}

\inferH{ml-faa}{}
{\hoare{\loc \mapsto n}{\fetchAdd{\loc}{k}}{\var.\, \var = n \wedge \loc \mapsto n + k}}

\inferH{ml-fork}
{P \Ra Q_0 * Q_1 \\\\ \hoare{Q_0}{\expr}{\TRUE} \and \hoare{Q_1}{\expr'}\propC}
{\hoare{P}{\fork\expr; \expr'}{\propC}[]}

\inferH{Ht-frame}
  {\hoare{\prop}{\expr}{\Ret\val.\propB}[]}
  {\hoare{\prop * \propC}{\expr}{\Ret\val.\propB * \propC}[]}

\inferH{Ht-csq}
  {\prop \vs \prop' \\
    \hoare{\prop'}{\expr}{\Ret\val.\propB'} \\   
   \All \val. \propB' \vs \propB}
  {\hoare{\prop}{\expr}{\Ret\val.\propB}}
\end{mathpar}
\caption{Selection of rules from Iris.}
\label{fig:hoarerules}
\end{figure}
}

\newcommand\invrulefig{\begin{figure}
    \begin{mathpar}
    \inferH{inv-alloc}
   {}
   {{\prop} \vs \Exists \iname. \knowInv{\iname}{\prop}}

    \inferH{inv-dup}
   {}
   {\knowInv{\iname}{\prop} \Ra \knowInv{\iname}{\prop} * \knowInv{\iname}{\prop}}

     \inferH{inv-open}
     {\text{(additional side conditions omitted)} \\ \hoare{\prop*\propC}{\expr}{\Ret\val.\prop*\propB} \and
       \physatomic{\expr}
     }
     {\hoare{\knowInv\iname\prop * \propC}{\expr}{\Ret\val.\propB}}

    \end{mathpar}
    \caption{Invariant rules.}
  \label{fig:invariantrules} 
  \end{figure}
}

\newcommand\counterresourcefig{\begin{figure}
  \begin{mathpar}
    \inferH{CountGeq}{}{\ownGhost{\gname}{\authFullCount{n}} *
             \ownGhost{\gname}{\authFragCount{q}{n'}} \Ra n \geq n'}
    
    \inferH{CountEq}{}{\ownGhost{\gname}{\authFullCount{n}}
              * \ownGhost{\gname}{\authFragCount{1}{n'}}
              \Ra n = n'}

    \inferH{CountPerm}{}
           {\ownGhost{\gname}{\authFragCount{q}{n}}
               * \ownGhost{\gname}{\authFragCount{q'}{n'}} \Ra q + q' \leq 1}

    \inferH{CountSep}{}
           {\ownGhost{\gname}{\authFragCount{q}{n}}
               * \ownGhost{\gname}{\authFragCount{q'}{n'}} \Lra
             \ownGhost{\gname}{\authFragCount{q+q'}{n + n'}}}

    \inferH{CountAlloc}{}{\TRUE \vs \exists \gname.\, \ownGhost{\gname}{\authFullCount{n}}
              * \ownGhost{\gname}{\authFragCount{1}{n}}}

    \inferH{CountUpd}{}
           {\ownGhost{\gname}{\authFullCount{(n+k)}}
             * \ownGhost{\gname}{\authFragCount{q}{n}}
             \vs \ownGhost{\gname}{\authFullCount{(n' + k)}}
             * \ownGhost{\gname}{\authFragCount{q}{n'}}}

  \end{mathpar}
\caption{Counter resource rules.}
\label{fig:counter-resource-rules}
\end{figure}
}

\newcommand\probrulesfig{\begin{figure}
    \begin{mathpar}
\inferH{Ht-Couple}
       {0 \leq n_1 / n_2 \leq 1 \\
         \ipCoupling{(\mret{\TRUE} \pchoice{\frac{n_1}{n_2}} \mret{\FALSE})}
                    {\pival}{\propC}
       }
       {\hoareHV{\ownProb{x \leftarrow \pival; F(x)}}{\flip{n_1, n_2}}{\Ret\val. \exists v'.\,
           \ownProb{F(v')} \wedge \propC(v, v')}}

\inferH{Ht-NonCouple}
       {0 \leq n_1 / n_2 \leq 1}
       {\hoare{\TRUE}{\flip{n_1, n_2}}{\Ret\val. (\val = \TRUE \vee \val = \FALSE)}}

\inferH{ProbLe}
       {\pival' \irrelleq \pival}
       {\ownProb{\pival} \Ra \ownProb{\pival'}}

\end{mathpar}

  \caption{Probabilistic rules.}
  \label{fig:prob-rules}
  \end{figure}
}
   
 \irisrulesfig
\counterresourcefig

Like other recent concurrency logics~\citep{vstbook, views, fcsl},
Iris lets users of the logic extend the notion of resource beyond just
heap fragments. These user-defined resources let us model the complex
protocols that govern how threads access shared state in a
concurrent system. Instead of describing the machinery that makes this
work, let us give an example of one of these user-defined resources:
the abstract ``counter'' resource.  There are two types of these
counter resources, represented by the following assertions:
\[ \ownGhost{\gname}{\authFullCount{n}} \qquad\text{and}\qquad \ownGhost{\gname}{\authFragCount{q}{n'}} \]
where $n$ and $n'$ are natural numbers, $0 < q \leq 1$ is a
rational number, and $\gname$ is an abstract name assigned to a particular
counter.  The $\authFullCount{n}$ resource represents a shared
counter that contains the value $n$. If we think of such a counter as
being composed of $n$ ``units'', then the resource
$\authFragCount{q}{n'}$ represents a ``stake'' or ownership of $n'$ of
the units in the global counter. The parameter $q$ is a fractional
permission~\citep{Boyland03} that lets us track how many threads
have such a stake; when $q = 1$, this represents full ownership,
so no other threads have a stake.\footnote{Note that the $q$
  is \emph{not} the fraction of the global counter value represented
  by the stake's value.}

Rules for using these assertions are given in
\figref{fig:counter-resource-rules}. The rules \ruleref{CountGeq} and \ruleref{CountEq} let us
conclude that the global counter value must be at least as big as any
stake's value; and when a stake's $q$ value is $1$, we furthermore
know that the counter and the stake value are the same.  The
rule \ruleref{CountSep} lets us join (or conversely, split) two stakes
by summing their permissions and their count values, subject to the (implicit)
constraint that $q$, $q'$, and $q + q'$ all lie in the interval $(0,
1]$. The \ruleref{CountAlloc} rule lets us create a new counter with
  some existentially quantified name; the $\vs$ connective here is
  a kind of implication in Iris which lets one modify or create a resource. Finally,
  $\ruleref{CountUpd}$ lets us modify a counter: if we own the global
  value and a stake, we can update the value and the stake, so long
  as we preserve the part of the counter value owned by other stakes (represented by $k$ in the rule).

\invrulefig

Of course, we need some way to connect these ``abstract'' resources to
the actual state of the program. The mechanism for doing this is an
\emph{invariant}. An invariant is an assertion that is dynamically
established at some point in the program, and then is guaranteed to
hold thereafter. We write $\knowInv\iname\prop$ for the assertion
which says that the invariant $\prop$ has been established with the
abstract name $\iname$.  If we have ownership of resources satisfying
$\prop$, we can use the rule \ruleref{inv-alloc} from \figref{fig:invariantrules} to establish $\prop$
as an invariant; we lose the resources and get back
$\knowInv\iname\prop$ with some fresh name $\iname$. If we know the
invariant $\knowInv\iname\prop$ holds and we are trying to prove some
Hoare triple about an expression $\expr$, we can use
\ruleref{inv-open} to ``open'' the invariant. This lets us add $\prop$
to the pre-condition of the triple we are trying to prove, but we need
to re-establish and give up $\prop$ in the post-condition in order to
``close'' the invariant. Moreover, to use this rule, $\expr$ must be
atomic: Since $\expr$ will reduce to a value in a single-step, this
ensures there is no intermediate step in which the invariant did not
hold. (In the statement of \ruleref{inv-open} we have omitted certain
side conditions that are used to ensure that the same invariant is not
opened multiple times simultaneously.)

For example, we can create the invariant $\knowInv{\iname}{\exists
  n.\, l \mapsto n * \ownGhost{\gname}{\authFullCount{n}}}$ to ensure
the physical heap location $l$ will always store the value represented
by the counter resource. Now a thread that owns a stake
$\ownGhost{\gname}{\authFragCount{q}{n'}}$ can read from $l$ or modify
it using a compare-and-swap by opening the invariant and updating the
global counter resource suitably with \ruleref{CountUpd}. This
resource and invariant pattern were used to verify a (non-approximate) concurrent counter
in the Iris Coq development. As we shall see, we are able to use
this same resource to verify the  approximate counter with our extensions.

\subsection{Probabilistic Rules}

We now describe how we extend Iris with probabilistic relational reasoning to obtain Polaris.  Our
goal is to be able to prove that there exists a suitable coupling between the
indexed valuation of a program and some set $\pival$ of indexed valuations.  The
existence of an appropriate coupling will let us use \thmref{thm:coupling}, so
that we can bound the expected values of our program by bounding the extrema of
$\pival$.

To motivate the rules of our extension, let us first give some more background on
how this kind of relational reasoning works in the pRHL logic of
\citeauthor{BartheEGHSS15}. The idea there, following \citeauthor{benton:popl04}'s
Relational Hoare Logic~\citep{benton:popl04}, is to replace Hoare triples with Hoare
\emph{quadruples}, which replace pre and post-conditions with pre and post-relations about \emph{pairs} of programs. Translated to our setting, we would have judgments of the form:
\[ \hoare{\prop}{\expr\,\,-\,\, \pival}{x, y.\, \propB} \]
where $\expr$ is the program we are trying to relate to $\pival$, and in the
post-relation $\propB$, we would substitute the return value of $\expr$ in for
$x$ and the return value of $\pival$ for $y$. Then, we would adapt the standard
Hoare rules to consider the pairs of steps of $\expr$ and $\pival$. Although the
work of \citeauthor{BartheEGHSS15} shows that this approach can be useful for
reasoning about non-concurrent probabilistic programs, there is an issue with
applying it in the concurrent setting: what do we do when $\expr$ forks a new
thread? We would then need to relate $\pival$ to multiple program expressions, and it's
not clear how to adapt \ruleref{ml-fork} to this quadruple style.

Instead, we adapt an idea originally developed by \citet{caresl} for non-probabilistic concurrent relational reasoning: rather than including the specification program $\pival$ as part of the Hoare judgment, we add
a new assertion $\ownProb{\pival}$ to the logic. That is, the specification computation becomes just another kind of ``resource'' that can be transferred between threads. 
We then add the probabilistic rules shown in
\figref{fig:prob-rules}. The rule \ruleref{Ht-Couple} lets us
establish a triple about a $\flip{n_1, n_2}$ command. The precondition
requires us to own a monadic computation of the form
$\mbind{x}{\pival}{F(x)}$, and we must exhibit a coupling between a
random choice between $\TRUE$ and $\FALSE$, (weighted by
$\frac{n_1}{n_2}$) and the monadic specification $\pival$. The post
condition says that we get back the monadic resource, but updated so
that it is now of the form $\ownProb{F(v')}$ for some $v'$. In
addition, $v$ (the outcome of the $\flip{n_1, n_2}$ command) and this
$v'$ are related by $R$, the postcondition of the coupling we
exhibited. This rule gives us a way to relate the execution of a concrete
expression $\expr$ to an execution of $\pival$.

Rule \ruleref{Ht-NonCouple} lets us handle a case where the concrete
program executes a probabilistic choice yet we do not want to relate this
to a step in the probabilistic specification. In this case, in the post-condition
we merely know that the return value was $\TRUE$ or $\FALSE$. 

Finally, $\ruleref{ProbLe}$ lets us replace our $\pival$ resource
with any $\pival'$ such that $\pival' \irrelleq \pival$. We use this to manipulate the $\pival$ into
a form that matches the precondition required by something like \ruleref{Ht-Couple}. Since $\pival' \equiv \pival$ implies $\pival' \irrelleq \pival$, it follows that if $\pival' \equiv \pival$ then $\ownProb{\pival} \Lra \ownProb{\pival'}$. 

Because $\ownProb{\pival}$ is just an assertion like any other, we can
control access to it between threads by storing it in an
invariant. This idea of representing a specification computation as a resource
assertion has been used in other separation logics based on Iris~\citep{iris-effects, TassarottiJH17}.

\probrulesfig

\subsection{Soundness}

The following soundness theorem for the logic will guarantee
that if we prove an appropriate triple involving $\ownProb{\pival}$,
the expected value of the concrete program will lie in the range
of the extrema of $\pival$: 

\begin{theorem}
  Let $\pival : \pivalmonad{T}$ for some type $T$,
  and let $f : \textdom{Val} \rightarrow \mathbb{R}$, $g : T \rightarrow \mathbb{R}$, and assume that $g$ is bounded on the support of $\pival$.
  Suppose
  \begin{equation*}
    \hoare{\ownProb{\pival}}{\expr}{v.\, \exists v'.\,
    \ownProb{\mret{v'}} \wedge f(v) = g(v')}
  \end{equation*}
  holds. Let $\sched$ be a scheduler
  such that $([\expr], \state)$ terminates in at
  most $n$ steps under $\sched$. Then $\exival{f}{\trstepivalN{\sched}{[\expr], \state}{n}}$
  exists and
  \[ \exMin{g}{\pival} \leq \exival{f}{\trstepivalN{\sched}{[\expr], \state}{n}}
  \leq \exMax{g}{\pival} \]
\label{thm:soundness}
\end{theorem}
This result only holds
for schedulers under which the program is guaranteed to terminate in
some number of steps; this is not that surprising, since the original
Iris is a partial correctness logic.

To prove this soundness theorem and validate the rules we have given,
we first change the definition of the Hoare triple in Iris so that if
the probabilistic resource is of the form
$\ownProb{\mbind{x}{\pival}{F(x)}}$ and the expression $\expr$ takes a
step, we must exhibit a coupling between $\expr$'s transition
(interpreted as an indexed valuation) and $\pival$. Then in the
soundness proof, as $\expr$ takes successive steps, we combine these
couplings together using \ruleref{Bind} from
\figref{fig:coupling-rules}; if $\expr$ terminates and the
post-condition matches the form stated in \thmref{thm:soundness}, then
we will have constructed a complete coupling between
$\trstepivalN{\sched}{[\expr], \state}{n}$ and the monadic
specification. Moreover, this will be an $R$-coupling with $R(x, y)
\eqdef f(x) = g(y)$. Hence, we can apply \thmref{thm:coupling} to
conclude the claim about the expected values.
\section{Example 1: Approximate Counters}
\label{sec:counters}
\newcommand{\LocInv}[2]{\textlog{LocInv}_{#1}(#2)}
\newcommand{\ProbInv}[2]{\textlog{ProbInv}_{#1, #2}}
\newcommand{\ACounter}[6]{\textlog{ACounter}_{#1, #2, #3}(#4, #5, #6)}
\newcommand{\ACounterNoArgs}{\textlog{ACounter}}
\newcommand{\gloc}{\gname_\textsf{l}}
\newcommand{\gpend}{\gname_\textsf{p}}
\newcommand{\gsum}{\gname_\textsf{c}}
\newcommand{\countTrue}{\textlog{countTrue}}
\newcommand{\foldLeft}{\textlog{foldLeft}}
\newcommand{\inarr}[1]{\begin{array}{@{}l@{}}#1\end{array}}
\newcommand{\lenTrue}[1]{\lvert#1\rvert_\textlog{t}}

\begin{figure}
\begin{mathpar}
  \inferH{ACounterNew}{}
           {\hoare{\ownProb{\approxN\ n\ 0}}{\alloc{0}}{\loc.\, \exists \gloc, \gpend, \gsum, \ACounter{\gloc}{\gpend}{\gsum}{\loc}{1}{n}}}

  \inferH{ACounterSep}{}{\ACounter{\gloc}{\gpend}{\gsum}{\loc}{q + q'}{n + n'} \Lra
             \ACounter{\gloc}{\gpend}{\gsum}{\loc}{q}{n} *
             \ACounter{\gloc}{\gpend}{\gsum}{\loc}{q'}{n'}}
  
  \inferH{ACounterIncr}{}
         {\hoare{\ACounter{\gloc}{\gpend}{\gsum}{\loc}{q}{n + 1}}{\incrfun\ \loc}{\ACounter{\gloc}{\gpend}{\gsum}{\loc}{q}{n}}}
         
  \inferH{ACounterRead}{}
         {\hoare{\ACounter{\gloc}{\gpend}{\gsum}{\loc}{1}{0}}{\readfun\ l}{v.\, \exists v'.\, \ownProb{\mret{v'}} \wedge v = v'}}
\end{mathpar}
  \caption{Specification for approximate counters.}
  \label{fig:counter-spec}
\end{figure}

\begin{figure}
     \[
     \begin{array}{ll}
       &\countTrue\ c\ lb\eqdef \foldLeft\ (\lambda\ \_\ b.\, \ifmatch{b}\ (\incrfun\ c)\ \elsematch\ ())\ lb\ () 
     \end{array}
     \]
   \vspace{1em}
 {\begin{center}
     $\hoareVC{\ownProb{\approxN\ (\lenTrue{lb_1} + \lenTrue{lb_2})\ 0}}
             {\bind{c}{\alloc{0}}{}}
             {\ACounter{\gloc}{\gpend}{\gsum}{c}{1}{\lenTrue{lb_1} + \lenTrue{lb_2}}}$
       \vspace{.5em}
 \begin{tabular}{@{}l||l@{}}
   \begin{tabular}{@{}l@{}}
     $\hoareVC{\ACounter{\gloc}{\gpend}{\gsum}{c}{1/2}{\lenTrue{lb_1}}}{\countTrue\ c\ lb_1}%
             {\ACounter{\gloc}{\gpend}{\gsum}{c}{1/2}{0}} $
   \end{tabular} &
   \begin{tabular}{@{}l@{}}
     $\hoareVC{\ACounter{\gloc}{\gpend}{\gsum}{c}{1/2}{\lenTrue{lb_2}}}{\countTrue\ c\ lb_2}%
             {\ACounter{\gloc}{\gpend}{\gsum}{c}{1/2}{0}} $
   \end{tabular}
 \end{tabular}
 $\hoareVC{\ACounter{\gloc}{\gpend}{\gsum}{c}{1}{0}}{\readfun\ c}{v.\,\exists v'.\, \ownProb{\mret{v'}} \wedge v = v'} $
 \end{center}}
  \caption{Example client using approximate counters.}
  \label{fig:counter-client}
\end{figure}

\begin{figure*}
  \begin{align*}
    \LocInv{\gloc}{\loc} &\eqdef \exists n.\, \loc \mapsto n * \ownGhost{\gloc}{\authFullCount{n}}
    \\ \\
    \ProbInv{\gpend}{\gsum}
    &\eqdef \exists n_1, n_2.\, \ownGhost{\gpend}{\authFullCount{n_1}} *
                               \ownGhost{\gsum}{\authFullCount{n_2}} *
                               (\ownProb{\approxN\ n_1\ n_2} \vee
                                \ownGhost{\gpend}{\authFragCount{1}{n_1}})
    \\ \\
    \ACounter{\gloc}{\gpend}{\gsum}{\loc}{q}{n} &\eqdef
    \exists \iname_1, \iname_2, n'.\,
    \knowInv{\iname_1}{\LocInv{\gloc}{l}} *
    \knowInv{\iname_2}{\ProbInv{\gpend}{\gsum}} *
    \ownGhost{\gloc}{\authFragCount{q}{n'}} *
    \ownGhost{\gpend}{\authFragCount{q}{n}} \\ &\qquad *
    \ownGhost{\gsum}{\authFragCount{q}{n'}}
  \end{align*}
  \caption{Invariants and definitions for proof.}
  \label{fig:counter-setup}
\end{figure*}
 
In this section we prove triples that relate the approximate counter
algorithm from \figref{fig:unbiased} to the monadic computation
$\approxN$ from \figref{fig:monad-counter}.

\subsection{Triples and Example Client}
The Hoare triples we have proved about this data structure are given in \figref{fig:counter-spec}.
The specification uses a predicate
$\ACounter{\gloc}{\gpend}{\gsum}{\loc}{q}{n}$, which can be treated by
a user as an abstract predicate representing the permission to perform
$n$ increments to the counter at $\loc$. The parameter $q$ is a
fractional permission that we use to track how many threads can access
the counter.  (Ignore the names $\gloc$, $\gpend$, and $\gsum$ -- we
will describe how they are used when we give the definition of
$\ACounterNoArgs$ later).  The triple \ruleref{ACounterNew} says that
we can create a new counter by allocating a reference cell containing
$0$. It takes the monadic specification $\ownProb{\approxN\ n\ 0}$ as
a precondition, and returns the full $\ACounterNoArgs$ permission
for $n$ increments.  The rule \ruleref{ACounterSep} lets us split or
join this $\ACounterNoArgs$ permission into pieces. If we have
permission to perform at least one increment, we can use
\ruleref{ACounterIncr}, which gives us back $\ACounterNoArgs$ with
permission to do one fewer increment. Finally, if we have
$\ACounterNoArgs$ with the full fractional permission 1, and there are
$0$ pending increments, we can use $\ruleref{ACounterRead}$. In the
post condition we get back $\ownProb{\mret \val}$, where $\val$ is the
value that the call to $\readfun$ returns.

At first this specification seems weak, but this is exactly what
we need for  \thmref{thm:soundness}. To see how we can use these triples to reason about a client program that uses the approximate counter, consider the example client in \figref{fig:counter-client}.
We start with a helper function
$\countTrue$, which takes an approximate counter $c$ and a list of booleans $lb$,
and counts the number of times $\TRUE$ occurs in $lb$ using the
counter. The client begins by creating a new counter $c$. It then runs
two threads in parallel that run $\countTrue$ on two lists $lb_1$ and
$lb_2$, using the shared counter $c$ -- we denote this parallel
composition using $\|$. The parent blocks until both threads finish
and then reads from the counter.\footnote{Of course, 
here the threads may as well maintain their own
exact counters and combine them at the end. But in a real
application such as \citep{TristanTS15}, there are
tens of millions of counters and hundreds of threads, so having each thread
maintain its own set of counters would be expensive.}

Refer to this client code as $\expr$.  If we write $\lenTrue{lb}$ for
the logical function giving the number of times $\TRUE$ occurs in
$lb$, then we would like to show that in expectation, $\expr$ returns
$\lenTrue{lb_1} + \lenTrue{lb_2}$. The derivation in
\figref{fig:counter-spec} shows that the triple
\[
\hoare{\ownProb{\approxN\ (\lenTrue{lb_1} + \lenTrue{lb_2})\ 0}}{\expr}{v.\,\exists v'.\, \ownProb{\mret{v'}} \wedge v = v'}
\]
holds. Moreover, it is not hard to show that for each $k$, there is an upper bound on the value returned by $\approxN\ k$, so by
\thmref{thm:soundness} we have:
\[ \exMin{\textlog{id}}{\approxN\ (\lenTrue{lb_1} + \lenTrue{lb_2})}
 \leq \exival{\textlog{id}}{\trstepivalN{\sched}{[\expr], \state}{n}}
 \]
 (and similarly for $\exMaxNoArgs$) for suitable $\sched$ and $n$. And, we have shown that
 $\exMin{\textlog{id}}{\approxN\ (\lenTrue{lb_1} + \lenTrue{lb_2})} = \lenTrue{lb_1} + \lenTrue{lb_2}$
 in \Sref{sec:monad-quant}, so we are done.

\subsection{Proofs of Triples}

The definition of $\ACounterNoArgs$ and the invariants used in the
proof are given in \figref{fig:counter-setup}. The proof uses three
counter resources to track: (1) the number of increments left to
perform in the monadic specification, (2) the accumulated count in the
monadic specification, and (3) the actual count currently stored in
the concrete program. We use two invariants to connect the counter
resources to these intended interpretations. First, we have
$\LocInv{\gloc}{l}$ which says that the counter resource named $\gloc$
stores some value $n$ and the physical location $l$ points to that
same value $n$. Then, assertion $\ProbInv{\gpend}{\gsum}$ says that
there are two counter resources containing some $n_1$ and $n_2$, and
the invariant either contains (a) the monadic specification resource
$\ownProb{\approxN\ n_1\ n_2}$ (\ie there are $n_1$ further increments
to perform, and the monadic counter has accumulated a value of $n_2$)
\emph{or} (b) it contains the complete stake for one of the counter
resources.  Then $\ACounterNoArgs$ says that these two invariants have
been set up with some names, and we own a stake in the $\gpend$
permission corresponding to the number of increments this permission
allows. Further, for some $n'$ there is a stake in the $\gloc$ and
$\gsum$ counters both equal to $n'$, which represents the total amount
that this permission has been used to add to the counter.

We will only describe the proofs of \ruleref{ACounterIncr} and \ruleref{ACounterRead},
since \ruleref{ACounterNew} is straight-forward.

\paragraph{Proof of \ruleref{ACounterIncr}.}
Eliminating the existentials in the definition of $\ACounterNoArgs$, we
get that the appropriate invariants have been set up and there is some
$n'$-stake in $\gloc$ and $\gsum$, along with the $n + 1$ stake in
$\gpend$.  The first step of $\incrfun\ l$ reads the value of
$l$; to perform this read the thread needs to own $l \mapsto \val$ for
some $\val$.  To get this resource, it opens the $\LocInv{\gloc}{\loc}$
invariant; after completing the read, the $l \mapsto \val$ resource is
returned to close the invariant.  The code then takes the minimum of
the value read and $\MAXCONST$, and binds this value to $k$.

It then performs $\flip{1, k+1}$. We want to use \ruleref{Ht-Couple} to
couple this flip with the monadic code. To do so, we first
open the invariant $\ProbInv{\gpend}{\gsum}$. We know this
will contain $\ownGhost{\gpend}{\authFullCount{n_1'}}$ and
$\ownGhost{\gsum}{\authFullCount{n_2'}}$ for some $n_1'$ and
$n_2'$, and either $\ownProb{\approxN\ n_1'\ n_2'}$ or a full stake
$\ownGhost{\gpend}{\authFragCount{1}{n_1'}}$. However, the latter is
impossible because the $\ACounter{\gloc}{\gpend}{\gsum}{\loc}{q}{n
 + 1}$ resource entails ownership of $\ownGhost{\gpend}{\authFragCount{q}{n + 1}}$,
but $q + 1 > 1$, contradicting \ruleref{CountPerm}. So, we obtain $\ownProb{\approxN\ n_1'\ n_2'}$. Now, by
\ruleref{CountGeq} we know that $n_1' \geq n + 1$, hence we can
unfold $\approxN\ n_1'\ n_2'$ to obtain
$\ownProb{\mbind{k}{\approxIncr}{\approxN\ (n_1' - 1)\ n_2'}}$.

We can now use $\ruleref{Ht-Couple}$ so long as we can exhibit a
coupling between the concrete program's coin flip and $\approxIncr$.
First, since $0 \leq k \leq \MAXCONST$, we can show that:
\begin{align*}
  &(\mret{k+1}) \pchoice{\frac{1}{k+1}} (\mret{0}) \\
  &\quad \pivalleq 
  (\mbind{x}{\mret 0 \ndchoice \cdots \ndchoice \mret \MAXCONST}
          {(\mret{x + 1} \pchoice{\frac{1}{x+1}} \mret{0})}) \\
  &\quad \equiv \approxIncr
\end{align*}
hence by \ruleref{Equiv}, it suffices to exhibit a coupling between
$(\mret{\TRUE} \pchoice{\frac{1}{k+1}} \mret{\FALSE})$ and
$(\mret{k+1} \pchoice{\frac{1}{k+1}} \mret{0})$. Take $R(x, y)$ to be
$(x = \TRUE \wedge y = k +1) \vee (x = \FALSE \wedge y =
0)$, then we can use \ruleref{P-Choice} and \ruleref{Ret} to prove the
existence of an $R$-coupling.

Applying \ruleref{Ht-Couple} with this coupling, we then have
$\ownProb{\approxN\ (n_1' - 1)\ (n_2' + v')}$ where $v'$ and the return value $v$
  of the $\flip{1, k}$ are related by $R$. We use \ruleref{CountUpd} to
  update the thread's stake in $\gpend$ resource to $n$, and the global
  value to $n_1' - 1$ (to record that a simulated increment has been performed),
  similarly, we update the thread's stake in the $\gsum$ counter to $n' + v'$
  and the global value to $n_2' + v'$ (to record the new total)
  and then close the $\ProbInv{\gpend}{\gsum}$ invariant.

  The code then cases on the value $v$ returned by the flip. If it is
  false, then $v'$ is $0$, the code returns, and the post condition
  holds. If $v$ is true, then $v' = k+1$, the amount that the code adds
  using a fetch-and-add. We therefore open the $\LocInv{\gloc}{l}$
  invariant again to get access to $l$, perform the increment and
  update the $\gloc$ counter and stake using \ruleref{CountUpd} to record the
  fact that we are adding $k+1$.

\paragraph{Proof of \ruleref{ACounterRead}}
The precondition $\ACounter{\gloc}{\gpend}{\gsum}{\loc}{1}{0}$
represents the full stake in each counter, and the $0$
argument means there are no pending increments to
perform. Thus, when we open the $\LocInv{\gloc}{\loc}$ and
$\ProbInv{\gpend}{\gsum}$ invariants we know that for some $n'$, $l \mapsto n'$ and
we have $\ownProb{\mret{n'}}$. So, we can read from $l$, knowing the
returned value will be $n'$. After reading, we must close the
invariant. This time we will keep the $\ownProb{\mret{n'}}$ resource
so that we can put it in the post condition, instead we give up
$\ownGhost{\gpend}{\authFragCount{1}{0}}$ to satisfy the disjunction
in $\ProbInv{\gpend}{\gsum}$.

\subsection{Variations}

In our mechanized proofs, we have verified two additional variations on this approximate
counter example. For the first variation, we consider a version of $\incrfun$ which
directly uses the current value it reads from the counter, rather than taking
the minimum of this value and $\MAXCONST$.

For the second variation, we address a limitation of the specification we have described so
far. Notice that to use the rules in \figref{fig:counter-spec} and obtain a
suitable triple to use with \thmref{thm:soundness}, the total number of calls to
$\incrfun$ must be a deterministic function of the program: we have to pick
some $n$ when we initialize the counter using \ruleref{ACounterNew}. In the case
of our example client, we chose $n$ to be the number of times that
$\textlog{true}$ occurred in the two lists. But what if the number of calls to
increment is itself probabilistic or non-deterministic? In this case we still
would like to know that the expected value returned by the approximate counter
is equal to the expected number of times the counter was incremented.  However,
if the number of times the counter is incremented is completely arbitrary, this
expected value may not exist! To guarantee that the expected value will exist,
our specification imposes an upper bound on the total number of increments that can be performed,
and then allows us to establish a coupling with the following
monadic computation:
\[
    \begin{array}{ll}
      &\approxN'\ 0\ t\ l \eqdef \mret{(t, l)} \\
      &\approxN'\ (n + 1)\ t\ l \eqdef 
      (\mret{(t, l)}) \ndchoice (\mbind{k}{\approxIncr}{} \approxN'\ n\ (t + 1)\ (l + k))
   \end{array}
\]
The first argument of $\approxN'$ gives an upper bound on the remaining number
of increments that can be performed, the second argument $t$ tracks the total
number of increments that have been done so far, and $l$ again tracks the
current value in the counter. In contrast to the original $\approxN$, before
performing each increment, there is a non-deterministic choice to simply return
$(t, l)$. Let $f$ be the function $\lambda (x, y).\, x - y$. We prove that
\[ \exMin{f}{\approxN'\ n\ 0\ 0} = \exMax{f}{\approxN'\ n\ 0\ 0} = 0 \]
\ie the expected value of the difference between the total number of increments and the value in the counter is $0$.
We have proved more flexible versions of the rules in \figref{fig:counter-spec} that use this
$\approxN'$ instead.
\section{Example 2: Concurrent Skip List}
\label{sec:skiplist}
\usetikzlibrary{calc,shapes.multipart,chains,arrows}

For our next example, we verify properties of a probabilistic concurrent skip
list. The code and proofs for this example are more complex, so for space
reasons we will give a high-level description of the algorithm and the triples
we have proved about it.

\subsection{Implementation}

\begin{figure}

\begin{tikzpicture}[list/.style={rectangle split, rectangle split parts=1, align=center, text width={width("$+\infty$")},
    draw, rectangle split horizontal}, >=stealth, start chain=1 going right, start chain=2 going right]
  \node[list,on chain=1] (LST) {$-\infty$};
  \node[list,on chain=1] (QT) {5};
  \node[list,on chain=1, draw=none, text opacity=0] (AT) {12};
  \node[list,on chain=1] (BT) {99};
  \node[list,on chain=1] (CT) {157};
  \node[list,on chain=1, draw=none, text opacity=0] (ZT) {167};
  \node[list,on chain=1] (RST) {$+\infty$};
  \node[list,on chain=2, below=.1in of LST] (LSB) {$-\infty$};
  \node[list,on chain=2] (QB) {5};
  \node[list,on chain=2] (AB) {12};
  \node[list,on chain=2] (BB) {99};
  \node[list,on chain=2] (CB) {157};
  \node[list,on chain=2] (ZB) {167};
  \node[list,on chain=2] (RSB) {$+\infty$};
  \draw[->] (LST) -- (QT);
  \draw[->] (QT) -- (BT);
  \draw[->] (BT) -- (CT);
  \draw[->] (CT) -- (RST);
  \draw[->] (LSB) -- (QB);
  \draw[->] (QB) -- (AB);
  \draw[->] (AB) -- (BB);
  \draw[->] (BB) -- (CB);
  \draw[->] (CB) -- (ZB);
  \draw[->] (ZB) -- (RSB);
  \draw[->] (LST) -- (LSB);
  \draw[->] (QT) -- (QB);
  \draw[->] (BT) -- (BB);
  \draw[->] (CT) -- (CB);
  \draw[->] (RST) -- (RSB);
\end{tikzpicture}
\caption{Diagram for 2-Level Concurrent Skip List}
\label{fig:skiplist}
\end{figure}
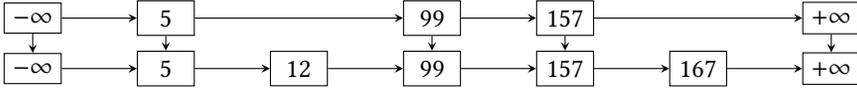

A skip list~\citep{Pugh90} is a data structure that can be used to
implement a dynamic set interface for ordered data. The implementation we
consider will only allow integer keys to be stored in the set.  The skip list
consists of several sorted linked lists, where the nodes in each list contain a
key. We visualize each list as running horizontally from left to right, with the
different lists stacked vertically above one another (see
\figref{fig:skiplist}). For simplicity we only allow 2 lists in our
implementation -- this still exposes most of the main concurrency issues
involved.  The set of keys contained in the top list is a subset of the keys
contained in the bottom list, and the node containing a key $k$ in the top list
includes a pointer to the corresponding node for $k$ in the list below it. At
the beginning and ends of each list, there are sentinel nodes containing the
minimum and maximum representable integer (which we write as $-\infty$ and
$+\infty$ in \figref{fig:skiplist}).

\paragraph{Non-concurrent implementation.} We first describe how operations on this data structure are implemented in the
sequential case. To check whether a key $k$ is contained in the set, we first
search for the key in the top list starting at the left sentinel. If we find a
node containing it, we return true. If not, we stop at the largest key $j < k$
in the list, and then follow the pointer in $j$'s node to the copy of $j$ in the
bottom list. We then resume searching for $k$ starting at node $j$ in the bottom
list. If $k$ is found in the bottom list we return true, otherwise the key is
not in the set so we return false. To insert $k$, we first find the nodes $N_t$
and $N_b$ with the largest keys $\leq k$ in the top and bottom list,
respectively. If we find that $k$ is already in either list, we stop and
return. Otherwise we execute $\flip{1/2}$. If it returns true, we insert new
nodes for key $k$ into both the top and bottom lists, after $N_t$ and
$N_b$. Otherwise, if it returns false, we only insert a node in the bottom list
after $N_b$. We call $N_t$ and $N_b$ the ``predecessor nodes'', because they
become the predecessors of $k$ if it is inserted into each list.

If we insert $n$ keys into the set, then in expectation $n/2$ of them will
appear in the top list. Then when searching for a key, we will be able to more
quickly descend down the top list, and either find they key there, or if not,
only have to examine a few additional nodes in the bottom list. Of course, it is
possible (though unlikely) that none or all of the nodes are inserted into the
top list, in which case we are effectively searching in a regular sorted linked
list. Later on, we will show how to use our program logic to derive a bound on
the expected number of comparisons needed to find a key in the list. We will not
handle deletion in our implementation, because if an adversarial client can
observe the state of the list, it can repeatedly delete and re-insert any key
that happens to end up in the top list, forcing the top list to be empty.

\paragraph{Adding concurrency.}

There are several ways to add support for concurrent operations to a skip
list. We will consider a simplified implementation inspired by that of
\citet{HLLS-Skip}. We add a lock to each node in the lists. Checking for whether
a key is in the set is the same as in the non-concurrent case, and no locks need
to be acquired.

To insert a key $k$, we again search for the predecessor nodes $N_t$ and $N_b$.
When we identify one of these nodes, we acquire its lock and then check that the
node after it has not changed in the time between when we examined its successor
and when the lock was acquired. If it has, that means another thread may have
inserted a new node with a larger key less than $k$, so we release the lock and
search for the predecessor again. Otherwise, so long as we hold the locks, we
are guaranteed that $N_t$ and $N_b$ will remain the proper predecessors for key
$k$. Having acquired both locks, we proceed as in the sequential case by generating a random bit, and on the basis of that bit we insert new nodes for $k$ into either both lists or just the bottom list. We then release the locks and return.

What effect does concurrency have on the number of nodes that must be examined
to find a key? None, so long as there are no concurrent insertions
happening while searching. The reason is that in the implementation we have just
described, the random choice is made \emph{after} acquiring the locks for
insertion. Thus, at the point the random choice is made, the ordering of
operations by threads cannot affect where the node will be inserted. 

However, to illustrate the subtleties involved, consider the following
variant. If we generate
the random bit \emph{before} acquiring locks for the predecessors, and the
resulting bit says we will only insert the node in the bottom list, then we only
need to acquire the lock for the bottom predecessor.
Now the distribution can be affected by the
scheduler. To see why, imagine two threads are trying to concurrently insert key
$k$ into the list. Suppose that the outcome of the first thread's random bit
generation indicates that it will insert the node only into the bottom list, but
the second thread will try to insert into both lists. Then the scheduler can
influence the distribution by pausing the second thread and letting the first
thread finish; when the second thread is eventually allowed to run, it will find
that $k$ is already in the list and so it will return without doing anything.
Although we have once more used adversarial language to describe the scheduler,
in this case such behavior could arise without having to
imagine any malice. Because the first thread only has to acquire a single lock,
it is plausible that it might tend to finish before the second thread. In the
following, we will just analyze the original version that we described.

\subsection{Monadic Model}

We follow the same pattern as in our verification of the approximate counter
example: we first define a monadic model of the data structure, bound
appropriate expected values of the monadic computation, and then describe
triples that can be used to prove the existence of a coupling between programs
using the skip list and the monadic model.

Our monadic model is the following:
\[
    \begin{array}{rcl}
      \skipPival\ \nil\ tl\ bl &\eqdef& \mret{(\sort(tl), \sort(bl))} \\
      \skipPival\ (k :: l)\ tl\ bl &\eqdef&
      \mbind{k'}{\bigcup\limits_{i \in k :: l} \mret{i}}{} \\
      &&\mbind{tl'}{(\mret{tl}) \pchoice{1/2} (\mret{k' :: tl})}{} \\
      &&\skipPival\ (\removeList\ k'\ (k :: l))\ tl'\ (k' :: bl){}
   \end{array}
\]
The computation $\skipPival\ l\ tl\ bl$ simulates adding keys from the list $l$
to a skip list, where the arguments $tl$ and $bl$ represent the keys in the top
and bottom lists of the skip list, respectively. If the first argument $l$ is
empty, it sorts $tl$ and $bl$ and returns the result. If $l$ is non-empty, it
first non-deterministically selects a key $k'$ from $l$. Then, with probability
1/2 it adds this key to $tl$. It then removes any copies of $k'$ from
$l$, and recurses to process the remaining elements with the updated top and
bottom lists. (There is no point in keeping the arguments $tl$ and $bl$ sorted
throughout the recursive calls in this monadic formulation.)

There are many quantitative properties of the skip list
that one might want to analyze (\eg the total number of nodes in both lists, the
probability that a large fraction of nodes lie only in the bottom list,
\emph{etc.}). As we alluded to above, we will bound the expected number of inequality
comparisons required to test for membership of a key in the skiplist. We define
a function $\skipcost{tl}{bl}{k}$ which gives the number of comparisons needed
to check if $k$ is in the skip list when the elements in the top and bottom lists
are $tl$ and $bl$, respectively:
\begin{align*}
  \topcost{tl}{k} &= 1 + |\{ i \in tl \ | \ \INTMIN < i < k \}| \\
  \rettop{tl}{k} &= \max(\{ i \in tl \ | i < k \} \cup \{ \INTMIN \}) \\
  \botcost{tl}{bl}{k} &= 1 + |\{ i \in bl \ | \ \rettop{tl}{k} < i < k \}| \\
  \skipcost{tl}{bl}{k} &=
  \begin{cases}
    \topcost{tl}{k} & \quad \text{if $k \in tl$} \\
    \topcost{tl}{k} + \botcost{tl}{bl}{k}  & \quad \text{if $k \not\in tl$}
  \end{cases}
\end{align*}
If the key $k$ is in the top list, then the number of comparisons is $1$ plus
the number of elements in the list less than $k$ ($\topcost{tl}{k}$).
If $k$ is not in the top list, then we must first still perform the
same number of comparisons while searching through the top list. Then
we search in the bottom list starting from the largest key less than $k$ that was in $tl$ ($\rettop{tl}{k}$). The total number of comparisons in the second list is the number
of keys between $\rettop{tl}{k}$ and $k$ ($\botcost{tl}{bl}{k})$.

We then bound $\exMax{\skipcostNoArg(-, -, k)}{\skipPival\ l\ \nil\ \nil}$, to
obtain an upper bound on the expected value of searching for a key $k$. Assuming $l$ has
no duplicates, the key $k$ and all keys in $l$ lie between $\INTMIN$ and $\INTMAX$, and there are $n$ keys less than $k$ in $l$, we show that:
\[ \exMax{\skipcostNoArg(-, -, k)}{\skipPival\ l\ \nil\ \nil} \leq
         1 + \frac{n}{2} + 2\left(1- \frac{1}{2^{n + 1}}\right) \]
This means that on average we have to do about half the number of comparisons
that would be required to search for the key in a regular sorted linked list.

\subsection{Triples}

\figref{fig:skip-spec} shows the triples we have proved about the skip list. The
specification defines an
assertion\linebreak$\Skipprop{\Gbundle}{q}{v}{S}{S_t}{S_b}$, which represents
permission to access a skip list whose top left sentinel is $v$. The argument
$\Gbundle$ is just a set of resource names (like the $\gamma$'s in the counter
example), $q$ is a fractional permission, $S$ is the finite set of keys which
may be added to the list, and $S_t$ and $S_b$ are a subset of the keys currently
in the top and bottom lists. Additional keys from $S$ may be in either $S_t$ or $S_b$,
but the owner of this permission assertion knows that they contain \emph{at least} these sets.

The expression $\newslfun$ creates a new skip list. The precondition for the
triple in \ruleref{SkipNew} requires us to own the monadic computation\footnote{
  Here, $S$ is a set, whereas the arguments to $\skipPival$ are lists. However,
  it is easy to show that if $l'$, $tl'$, and $bl'$ are permutations of the
  lists $l$, $tl$, and $bl$, respectively, then $\skipPival\ l\ tl\ bl \equiv
  \skipPival\ l'\ tl'\ bl'$, so it makes no difference if we treat the first
  argument instead as an unordered set.}
$\ownProb{\skipPival\ S\ \nil\ \nil}$. The post condition gives the full
permission ($q = 1$) to access the skip list, with empty top and bottom lists.
To use this rule, all the keys in $S$ must be between $\INTMIN$ and $\INTMAX$.
Notice here that the set of keys $S$ which will be added to the skiplist must be
deterministic, so that it can be decided in this precondition (much like our
original specification for the approximate counters required the total number of
increments to be deterministic).  This restriction is important: if the keys to
be added are non-deterministically selected, and a client can observe the state
of the skip list, it can insert a special sequence of keys in such a way so as
to force a large number of comparisons to find a particular target key.

We use $\addfun$ to insert a key $k$ into the skip list. The post condition in
$\ruleref{SkipAdd}$ indicates that we now know that the added key $k$ is in the
bottom list. On the other hand, the client does not know whether the key was
added to the top list or not, so in the permission for the post condition, the
contents of the top list are given by some existentially quantified $S_t'$.

The function $\memfun$ checks whether a key is in the skip list. It returns a
pair $(b, z)$, where $b$ is a boolean indicating whether the key was in the set
or not, and $z$ is the number of key comparisons performed. The triple
\ruleref{SkipMem} says that if we have the full permission
for the skip list, then the boolean $b$ indeed
reflects whether the key is in the set or not, and $z$ is in fact equal to the
cost function $\skipcost{S_t}{S_b}{k}$ we defined above.
(One can also prove triples for when we have less than
the full permission of the list (\ie $q < 1$), but we have not done so.)

The rule \ruleref{SkipSep} lets us split and join together the $\SkippropNoArgs$
permission so that separate threads can use the skip list. Finally,
\ruleref{SkipShift} lets us do a view shift to convert a full
$\SkippropNoArgs$ permission in which we have added all the keys in $S$
to the skip list back into a $\ownProbNoArgs$ permission in which
the monadic computation is finished.
This lets us prove triples of the form required by \thmref{thm:soundness}.

In our mechanized proofs we have used this specification to verify a simple client in which
two threads concurrently add lists of integers to a skip list set, and then after they both
finish, one looks up a key using $\memfun$ and returns the number of comparisons performed.

\begin{figure}
\begin{mathpar}
  \inferH{SkipNew}{\forall k \in S.\, \INTMIN < k < \INTMAX}
           {\hoare{\ownProb{\skipPival\ S\ \nil\ \nil}}{\newslfun}{v.\, \exists \Gbundle. \, \Skipprop{\Gbundle}{1}{v}{S}{\emptyset}{\emptyset}}}

  \inferH{SkipAdd}{k \in S}
           {\hoare{\Skipprop{\Gbundle}{q}{v}{S}{S_t}{S_b}}{\addfun\ v\ k}{\exists S_t'. \, \Skipprop{\Gbundle}{q}{v}{S}{S_t'}{S_b \cup \{k\}}}}

  \inferH{SkipMem}{\INTMIN < k < \INTMAX}
         {\hoareHV{\Skipprop{\Gbundle}{1}{v}{S}{S_t}{S_b}}{\memfun\ v\ k}
           {(b, z). \begin{aligned}&\Skipprop{\Gbundle}{1}{v}{S}{S_t}{S_b} * (b = \TRUE \Ra k \in S_b)\\& * (b = \FALSE \Ra k \notin S_t \cup S_b) * (z = \skipcost{S_t}{S_b}{k})\end{aligned}}}

  \inferH{SkipSep}{}
    {\Skipprop{\Gbundle}{q + q'}{v}{S}{S_t \cup S_t'}{S_b \cup S_b'} \Lra
      \Skipprop{\Gbundle}{q}{v}{S}{S_t}{S_b} *
      \Skipprop{\Gbundle}{q'}{v}{S}{S_t'}{S_b'}}

    \inferH{SkipShift}{}
    {\Skipprop{\Gbundle}{1}{v}{S}{S_t}{S} \vs \ownProb{\mret (\sort(S_t), \sort(S))}}
\end{mathpar}
  \caption{Specification for skip list.}
  \label{fig:skip-spec}
\end{figure}

\section{Related Work}
\label{sec:related}

As we have described, the key to our approach has been to synthesize
ideas from several lines of related work. We now mention
further related work.

\paragraph{Probabilistic logics.}

\citet{McIverRS16} present a probabilistic version of rely-guarantee
logic~\citep{rg}.
Like the original rely-guarantee logic, this logic does not permit
local reasoning: one must check stability against rely-guarantee conditions that
refer to the global state of the program. More recent concurrency logics have
combined rely-guarantee reasoning with the local reasoning features of
concurrent separation logic~\citep{rgsep, lrg}. Iris, and our extensions,
incorporate these ideas, which is what enables us to give specifications like
those in \figref{fig:skip-spec} -- clients can use the skip list without having
to reason about interference involving the underlying state of the list.
Since many of the interesting uses of randomization in the concurrent setting
are in implementations of \emph{data structures}, it is important to be able
to provide these more abstract specifications.

Recently, \citet{QuantSepLog} have developed a version of (non-concurrent)
separation logic for reasoning about sequential probabilistic programs with dynamic memory
allocation. They verify an example of a program which probabilistically appends
nodes to a list (so that the length of the list is geometrically distributed),
and a tree deletion procedure which only probabilistically deletes nodes. Instead
of using relational reasoning, assertions in their logic denote probabilities/expected
values, and rules are given for computing and bounding these probabilities.

Several program logics for probabilistic reasoning (\eg \citep{MorganMS96}) are
designed to reason about languages that have primitives for both
probabilistic choice and (demonic) non-deterministic choice.  However,
in that work, non-determinism was not used for modelling concurrency,
but rather a program which might be ``underspecified'' and have
multiple possible implementations of a component, of which one is
selected non-deterministically, as in \citeauthor{Dijkstra75}'s~\citep{Dijkstra75} work on
GCL.

\citet{BartheEGHSS15}
were the first to connect the idea of coupling to the kind of
probabilistic relational reasoning done in pRHL, an earlier logic
by \citet{BartheGB12}. Since then, different results from the theory
of coupling and variants of couplings have been used to extend
pRHL~\citep{BartheEGHS17, BartheGHS17, BartheEHSS17, HsuThesis}.

Iris uses a step-indexed semantic model to support impredicative features of the
logic, but we do not make special use of step-indexing in our
extensions. However, \citet{GuardedMarkov} have shown that a kind of
step-indexed model can be used to reason about more general kinds of couplings
(so-called ``shift couplings''). Previously, \citet{BizjakB15} developed a
step-indexed logical relation for a higher-order language with random choice.

\paragraph{Denotational semantics.} A number of denotational models
combining probabilistic and non-deterministic choice have been
developed~\citep{Jones90, VaraccaW06, TixKP09a, Varacca02, Mislove00,
  Goubault-Larrecq15}. Our soundness theorem considers a scheduler
which \emph{deterministically} selects which thread to run next.
\citet{VaraccaW06} showed that their monadic encoding, which we have
used in our work, gives an adequate semantic model for an imperative
language with this kind of deterministic scheduler.  An alternative is
to permit the scheduler to also make random choices when selecting
which thread to run. \mbox{\citeauthor{VaraccaW06}} show that in this case,
an alternative monad developed by \citet{Mislove00} and
\citet{TixKP09a} gives an adequate model. It would be interesting to
use this latter monad in our program logic to reason about behavior
under probabilistic schedulers.

\paragraph{Linearizability and quiescent consistency.}
For non-randomized concurrent data structures, one important correctness
criterion is \emph{linearizability}~\citep{linearizability}, which ensures that
we can treat the execution of operations on the data structure as if they
happened in atomic steps. Many program logics have been developed for
establishing linearizability itself or related notions of
atomicity~\citep{caresl, tada, reloc, vafeiadis-thesis}. However,
\citet{GolabHW11} show that when clients of data structures can make randomized
choices, linearizable implementations can indeed be distinguished from versions
that are truly atomic. They propose an alternative condition called \emph{strong
  linearizability} which resolves this issue, but they do not consider
randomized implementations of data structures, only randomized clients.

The bounds we have proved 
only apply to ``stable'' states of the data structure in which there are no on-going
 modifications.
In the analysis of
non-probabilistic concurrent data structures, there is a condition
weaker than linearizability known as \emph{quiescent consistency}, whose definition considers
groups of operations separated by gaps of time in which no modifications occur.
\citet{SergeyNBD16} have developed ways of
reasoning about quiescent consistent data structures in a separation logic, and
it might be possible to adapt their approach to the probabilistic setting.

\section{Conclusion}
\label{sec:conclusion}
We have developed a concurrent program logic that can be used for
probabilistic relational reasoning, and have used it to verify two
realistic examples of randomized concurrent algorithms.
Moreover, we have mechanized all the results described here in Coq by
modifying the prior Coq formalization of Iris.
The development is available at \url{https://github.com/jtassarotti/polaris}.

\begin{acks}

The authors thank Jean-Baptiste Tristan, Jan Hoffmann, Derek Dreyer,
Guy L. Steele, Victor Luchangco, Jeremy Avigad, Jon Sterling, Justin Hsu, and Daniel Gratzer for
feedback and discussions related to this work.

This work was supported by a gift from Oracle Labs.  This
research was conducted with U.S. Government support under and awarded
by DoD, Air Force Office of Scientific Research, National Defense
Science and Engineering Graduate (NDSEG) Fellowship, 32 CFR 168a.  Any
opinions, findings and conclusions or recommendations expressed in
this material are those of the authors and do not necessarily reflect
the views of these organizations.

\end{acks}

\bibliography{bib}

\ifdefined\INCLUDEAPPENDIX
 \appendix
 \section{Appendix}
\label{appendix}

In this appendix, we describe in more detail how we modify the
definition of Hoare triples in Iris in order to support our
probabilistic extensions. Recall from the description in the paper
that the main idea is to augment the definition of Hoare triples so
that a derivation of a Hoare triple will encode a coupling between
each step of the concrete program $\expr$ and the monadic
specification.

\newcommand{\pinterp}[1]{\textlog{PInterp}(#1)}
\newcommand{\primStep}{\textlog{primStep}}
\newcommand{\prstepival}[2]{\textlog{primStep}(#1, #2)}

From here on, we assume familiarity with the Iris model.
Let $\textdom{ProbState}$ be the type
$\Sigma_{T: \textdom{Type}}\,\nondetmonad{\ivalmonad{T}}$. Given two
terms $(T_1, \pival_1)$ and $(T_2, \pival_2)$ of type
$\textdom{ProbState}$, we say $(T_1, \pival_1) \equiv (T_2, \pival_2)$
if $T_1 = T_2$ and $\pival_1 \equiv \pival_2$.  Using this equivalence
relation, we impose a discrete OFE structure on $\textdom{ProbState}$.
Here, we will often simply omit the type $T$ when writing an element
of $\textdom{ProbState}$.

The monad specification resource is handled much in the same way that
physical state is in Iris.  We represent the monoid specification code
using the ``authoritative exclusive resource'' construction. Define:
\begin{mathpar}
\pinterp{\pival} \eqdef \authfull{\exm\ \pival}

\ownProb{\pival} \eqdef \exists\pival'.\ \pival \irrelleq \pival' * \authfrag{\exinj(\pival')}
\end{mathpar}

In Iris, Hoare triples are defined in terms of weakest-preconditions,
so we actually need to modify the definition of the latter.  Recall
the following definition of weakest-precondition in Iris 3.0~\citep{IrisManual}, which is
indexed by a state interpretation function $S$.
\newcommand{\stateinterp}{S}
\newcommand{\metanone}{\textlog{None}}
\newcommand{\metasome}{\textlog{Some}}
\newcommand{\pure}[1]{#1}
\newcommand\vsWand{{\displaystyle\equiv\kern-1.6ex-\kern-1.5ex\smash{\scalerel*{\vphantom-\ast}{\sum}}\kern-0.2ex}}
\NewDocumentCommand \vsW {O{} O{}} {\vsGen[#1]{\vsWand}[#2]}
\begin{align*}
  \textlog{wp}^\stateinterp \eqdef{}& \MU \textdom{wp}. \Lam \mask, \expr, \pred. \\
        & (\Exists\val. \toval(\expr) = \val \land \pvs_\mask \pred(\val)) \lor {}\\
        & \Bigl(\toval(\expr) = \metanone \land \All \state. \stateinterp(\state) \wand{}\\
        &\qquad \pvs[\mask][\emptyset]\Bigl(\red(\expr, \state) * \later\All \expr', \state', \tpool. (\expr, \state \step \expr', \state', \tpool) \wand {}\\
        &\qquad\qquad \pvs[\emptyset][\mask]\Bigl(\stateinterp(\state') * \textdom{wp}(\mask, \expr', \pred) * \Sep_{\expr'' \in \tpool} \textdom{wp}(\top, \expr'', \Lam \any. \TRUE)\Bigr)\Bigr)\Bigr) 
\end{align*}
The definition is a guarded fixed-point, composed of a disjunction
which handles two cases: (1) either the expression $\expr$ is value,
in which case the post-condition $\pred$ should hold for that value,
or (2) it is not a value, in which case for each possible state
$\state$, given the interpretation of $\state$, we need to show
$\expr$ is reducible, and then recursively show that for each thing
which $\expr$ could step to, we will be able to update the state
interpretation appropriately and recursively prove
weakest-precondition for the reduct.

We write $\primStep(\expr, \state)$ for the indexed valuation of type $\textdom{Option}(\textdom{Expr}\ \times\ \textdom{State}\ \times\ \textdom{List Expr})$ which returns reducts of $\expr; \state$ or $\textlog{None}$ if $\expr; \state$ is not reducible.

In our version, the definition of weakest precondition becomes:
\begin{align*}
  \textlog{wp}^\stateinterp \eqdef{}& \MU \textdom{wp}. \Lam \mask, \expr, \pred. \\
        & (\Exists\val. \toval(\expr) = \val \land \pvs_\mask \pred(\val)) \lor {}\\
        & \Bigl(\toval(\expr) = \metanone \land \All \state, \isnew{\pival}. (\stateinterp(\state) * \isnew{\pinterp{\pival}}) \wand  {}\\
        &\qquad \pvs[\mask][\emptyset]\Bigl(\pure{\red(\expr, \state)} * \later \isnew{\Exists R, \pival', F.}  \\
        &\qquad\qquad\quad \isnew{\pure{(\mbind{x}{\pival'}{F(x)} \irrelleq \pival) \land \ipCoupling{\prstepival{\expr}{\state}}{\pival'}{R}}} \land {} \All \expr', \state', \tpool, \isnew{x}.  \\
        &\qquad\qquad\quad \pure{(\expr, \state \step \expr', \state', \tpool) \land \isnew{R(\textlog{Some}(\expr_2, \state_2, \tpool), x)}} \wand {}\\
        &\qquad\qquad\quad \pvs[\emptyset][\mask]\Bigl(\stateinterp(\state') * \isnew{\pinterp{F(x)}} * \textdom{wp}(\mask, \expr', \pred) * {} \\
        &\qquad\qquad\quad\qquad \Sep_{\expr'' \in \tpool} \textdom{wp}(\top, \expr'', \Lam \any. \TRUE)\Bigr)\Bigr)\Bigr) \\
\end{align*}
Compared to the original definition of weakest precondition, this says that in
the case where $\expr$ is not a value, we get not only the state interpretation
$\stateinterp(\state)$ for some state, but also the authoritative resource
version $\pinterp{\pival}$ of some monadic computation. From there, we have to
show that $\pival$ is greater than or equal to $\mbind{x}{\pival'}{F(x)}$ under the $\irrelleq$ ordering for
some $\pival'$ and $F$, and must exhibit an $R$-coupling
between the reduction of $\expr$ in state $\state$ and $\pival'$.
Then, as in the lifting lemma, in addition to quantifying over what $\expr$ can reduce to, we also quantify
over values $x$ that can be returned by $\pival'$, and can assume that
the reducts of $\expr$ and $x$ are related by $R$. Besides establishing the new state interpretation and weakest preconditions
for the results of stepping $\expr$, we also have to update the authoritative copy of the monadic computation to $\pinterp{F(x)}$.
 \fi

\end{document}